\documentclass[review]{elsarticle}

\usepackage{amsthm}
\usepackage{subfig}
\usepackage[table,x11names]{xcolor}
\usepackage{amsmath}
\usepackage{transparent}

\usepackage{smartdiagram}
\usepackage{graphicx}
\usepackage{soul}
\usepackage{float}
\usepackage{multirow}
\usepackage{booktabs} 
\usepackage{etoolbox} 
\usepackage[vlined,boxed,commentsnumbered, ruled, linesnumbered]{algorithm2e}
\SetKw{KwBy}{by}
\usepackage{tablefootnote}
\makeatletter
\newcommand{\removelatexerror}{\let\@latex@error\@gobble}
\makeatother

\usepackage{smartdiagram}
\usepackage[margin=3cm]{geometry}

\usepackage{footnote}

\usepackage[margin=3cm]{geometry}

\makeatletter
\def\BState{\State\hskip-\ALG@thistlm}
\makeatother

\usepackage{setspace}

\usepackage{breakcites}
\usepackage{enumerate}
\usepackage{makecell}

\usepackage{float}
\usepackage{listings}
\usepackage{amssymb} 
\usepackage{tikz}
\usetikzlibrary{trees}
\usepackage{paralist}
\usepackage{ragged2e}
\usepackage[framemethod=tikz]{mdframed}

\usetikzlibrary{positioning,arrows.meta}

\definecolor{arrowblue}{RGB}{98,145,224}

\usepackage{soul}

\usepackage[colorinlistoftodos]{todonotes}

\usepackage{lineno,hyperref}
\usepackage{cleveref}
\usepackage{setspace}
\modulolinenumbers[1]

\journal{Preprint}

\begin{document}

\begin{frontmatter}

\title{PPaaS: Privacy Preservation as a Service}

\author[mymainaddress,mysecondaryaddress]{M.A.P.~Chamikara
	\corref{mycorrespondingauthor}}
\cortext[mycorrespondingauthor]{Corresponding author}
\ead{pathumchamikara.mahawagaarachchige@rmit.edu.au}

\author[mymainaddress]{P.~Bertok}
\author[mymainaddress]{I.~Khalil}
\author[mysecondaryaddress]{D.~Liu}
\author[mysecondaryaddress]{S.~Camtepe}

\address[mymainaddress]{RMIT University, Australia}
\address[mysecondaryaddress]{CSIRO Data61, Australia}

\begin{abstract}
\begin{mdframed}[backgroundcolor=green!50,rightline=false,leftline=false]
\centering 
The published article can be found at \url{https://doi.org/10.1016/j.comcom.2021.04.006}
\end{mdframed}

Personally identifiable information (PII) can find its way into cyberspace through various channels, and many potential sources can leak such information. Data sharing (e.g. cross-agency data sharing) for machine learning and analytics is one of the important components in data science. However, due to privacy concerns, data should be enforced with strong privacy guarantees before sharing. Different privacy-preserving approaches were developed for privacy preserving data sharing; however, identifying the best privacy-preservation approach for the privacy-preservation of a certain dataset is still a challenge. Different parameters can influence the efficacy of the process, such as the characteristics of the input dataset, the strength of the privacy-preservation approach, and the expected level of utility of the resulting dataset (on the corresponding data mining application such as classification).  This paper presents a framework named  \underline{P}rivacy \underline{P}reservation \underline{a}s \underline{a} \underline{S}ervice (PPaaS) to reduce this complexity.  The proposed method employs selective privacy preservation via data perturbation and looks at different dynamics that can influence the quality of the privacy preservation of a dataset. 
PPaaS includes pools of data perturbation methods, and for each application and the input dataset, PPaaS selects the most suitable data perturbation approach after rigorous evaluation. It enhances the usability of privacy-preserving methods within its pool; it is a generic platform that can be used to sanitize big data in a granular, application-specific manner by employing a suitable combination of diverse privacy-preserving algorithms to provide a proper balance between privacy and utility. 
\end{abstract}

\end{frontmatter}

%\linenumbers
\section{Introduction}
\label{introsection}

Cyberspace users cannot easily avoid the possibility of their identity being incorporated in data that exposes various aspects of their lives~\cite{chamikara2018efficient}. Our day-to-day life activities are tracked by smart devices, and the unavoidable exposure of personally identifiable information (PII) such as fingerprint, facial features can lead to massive privacy loss. 
The heavy use of PII in social networks, in the health-care industry, and by insurance companies, in smart grids makes privacy protection of PII extremely complex.  Literature shows more than a few methods to address the growing concerns related to user privacy. Among these methods, disclosure control of microdata has become widely popular in the domain of data mining~\cite{chamikara2018efficient}; it works by applying different privacy-preserving mechanisms to the data before releasing them for analysis. Privacy-preserving data mining (PPDM) applies disclosure control to data mining in order to preserve privacy while generating knowledge~\cite{chamikara2018efficient}.

The main approaches to PPDM use data perturbation (modification) or encryption; literature shows a plethora of privacy preservation approaches under these two categories~\cite{chamikara2019efficient}. There has been more interest in data perturbation due to its lower complexity compared to encryption. Additive perturbation, random rotation, geometric perturbation, randomized response, random projection, microaggregation, hybrid perturbation, data condensation, data wrapping, data rounding, and data swapping are some examples of basic data perturbation algorithms, which show different behavior on different applications and datasets~\cite{torra2017fuzzy, hasan2016effective,aldeen2015comprehensive,okkalioglu2015survey, dwork2014algorithmic}. We can also find a number of hybrid approaches that combine basic perturbation approaches.

\begin{figure}[H]
	\centering
	\scalebox{0.7}{
	\transparent{0.8}\includegraphics[width=0.9\textwidth, trim=0.5cm 0cm 0.5cm
	 0cm]{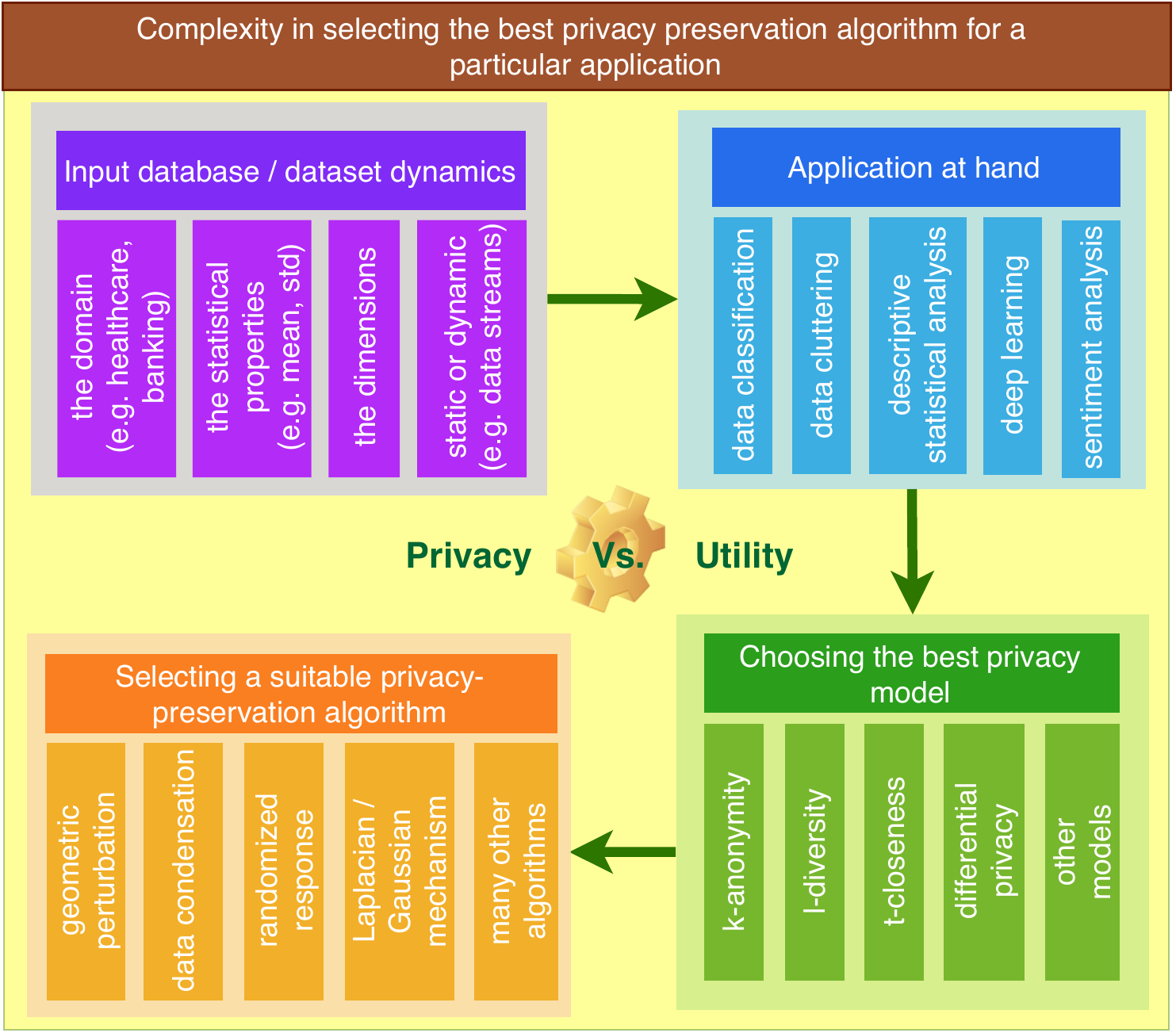}	 }
	\caption{Complexity of selecting the best privacy preservation approach for a particular application/ database}
	\label{sysel}
\end{figure}

The availability of many privacy preservation approaches has its drawback: the selection of the optimal perturbation algorithm for a particular problem can be quite complex; Figure \ref{sysel} shows different constraints that need to be considered. Different characteristics of privacy models (e.g. k-anonymity, l-diversity, t-closeness, differential privacy~(\cite{chamikara2019efficient})),  different properties of privacy preservation algorithms (e.g. geometric perturbation, data condensation, randomized response), different dynamics of the input data (e.g. the statistical properties, the dimensions), and different types of applications at hand (e.g. data clustering, deep learning) are examples of the attributes that influence the effectiveness of privacy preservation and the usability of the results. 
At the same time, this diversity enables the selection of the privacy preservation algorithm that best suits a particular application. 
There is no generic approach to identify the exact levels of privacy loss vs. utility loss, given a list of privacy preservation algorithms on specific applications and datasets. Furthermore, many privacy preservation approaches fall out of favour because their applicability is not properly identified. We introduce a new approach named ``Privacy Preservation as a Service " (PPaaS) that employs a novel strategy to apply customized perturbation based on the requirements of the problem at hand and the characteristics of the input dataset.

PPaaS presents a unified service that understands data requesters' needs and data owners' (who have full access privileges to the raw input databases which are represented by the lowest layer Figure \ref{framework}) requirements; it can facilitate privacy-preserving data sharing and can identify the best data perturbation approach. While an exhaustive analysis of all privacy protection methods for a given set of data is not feasible, a quantitative evaluation of selected relevant methods can significantly improve the efficacy of privacy protection. An appropriate set of performance and security metrics describes the quality of such a service, which is used to tailor the best privacy preservation to stakeholders' needs. The proposed framework collects efficient privacy preservation methods into a pool and applies the approach that best suits both data owner and data requester to the data before making the data available. The selection of the best perturbed dataset is done based on attack resistance analysis integrated into privacy and utility evaluation using a fuzzy inference system (FIS).

\subsection{Rationale and technical novelty}
Developing generic privacy-preserving methods for data mining and statistics is challenging due to the large number of constraints that need to be considered. 
As the complexity of the applications increases, generic approaches often end up with low utility or low privacy~(\cite{arachchige2019local}). Many researchers try to overcome this by focusing on a distinct objective (e.g privacy in deep learning)~(\cite{abadi2016deep,shokri2015privacy}). 
As a result, there are a number of algorithms for some areas such as deep learning, with many viable privacy preservation solutions~(\cite{zhao2019differential}). The algorithms having unique features and characteristics, choosing the best one for a particular case can be highly complex. 

PPaaS reduces the burden of choosing the optimal privacy-preserving algorithm and providing the best protection for the application and dataset at hand by introducing a unified service for the purpose.
Since there can be more than one method appropriate for a particular application and dataset, empirical evaluation is utilized in this process. 
PPaaS manages a pool of data perturbation algorithms suitable for particular applications and a pool of potential data reconstruction attacks that can reconstruct the original data from perturbed data. When a certain application/dataset is presented, PPaaS assesses the data perturbation algorithms and produces a unified metric named fuzzy index (FI) derived from a fuzzy model. Conventionally, a fuzzy model is used to model the vagueness and impreciseness of information in a real-world problem using fuzzy sets. In PPaaS, we use a fuzzy model to select the best perturbed dataset based on privacy, attack resistance, and utility from the corresponding perturbed instances. PPaaS utilizes only robust privacy preservation approaches for data perturbation in order to avoid potential data reconstruction attacks on the perturbed data. Privacy protection is aiming at reducing the leakage of information in responses to legitimate queries. However, it was shown that the existing privacy preservation approaches are still vulnerable to different privacy attacks~\cite{zigomitros2020survey}. The attack resistance module of PPaaS is responsible for evaluating the robustness of a particular perturbed dataset against different attacks. We measure the robustness/resistance of a particular perturbed dataset against data reconstruction attacks and generate a minimum guarantee or resistance (MGR); the higher the miminum guarantee, the better the perturbation algorithm used for the perturbation of the dataset. We use quantitative definitions of utility and privacy along with MGR  as inputs to the Fuzzy model. The higher the fuzzy index, the better the balance between privacy and utility under the given circumstances. The release of a particular output depends on a configurable threshold value of the corresponding FI. If the required threshold is not reached, the application of the corresponding pool is assessed repeatedly with different algorithms and parameters until one of the privacy preservation algorithms in the pool generates a satisfactory FI ($\geq$ threshold $FI$) for an application and dataset or the user-defined maximum number of iterations reached.   With this approach, users are guaranteed to be given the best possible privacy preservation while providing optimal utility.

\section{Literature}
\label{litrev}
Data privacy focuses on impeding the estimation of the original data from the sanitized data, while utility concentrates on preserving application-specific properties and information~(\cite{aggarwal2015privacy}). 
It has been noted that privacy preservation mechanisms decrease utility in general, i.e. they reduce utility to improve privacy, and finding a trade-off between privacy protection and data utility is an important issue ~(\cite{xu2015privacy}).
In fact, privacy and utility are often conflicting requirements: privacy-preserving algorithms provide privacy at the expense of utility.
Privacy is often preserved by modifying or perturbing the original data, and a common way of measuring the utility of a privacy-preserving method is to investigate perturbation biases ~(\cite{wilson2008protecting}). This bias is the difference between the result of a query on the perturbed data and the result of the same query on the original data. Wilson et al. examined different data perturbation methods and identified Type A, B, C, and D biases, along with an additional bias named Data Mining (DM) bias~(\cite{wilson2008protecting}). Type A bias occurs when the perturbation of a given attribute causes summary measures to change. Type B bias is the result of the perturbation changing the relationships between confidential attributes, while in case of Type C bias, the relationship between confidential and non-confidential attributes changes. Type D bias means that the underlying distribution of the data was affected by the sanitization process. If Type DM bias exists, data mining tools will perform less accurately on the perturbed data than they would on the original dataset.  

An investigation of existing privacy preservation approaches also suggests that they often suffer from utility or privacy issues when they are considered for generic applications~(\cite{chamikara2019efficient}).
Methods such as additive perturbation with noise (for differentially private data) can produce low utility due to the highly randomized nature of added noise~(\cite{agrawal2000privacy,okkalioglu2015survey}). Randomized response, another privacy preservation approach, has the same issue and produces low utility data due to high randomization~(\cite{dwork2014algorithmic}). Methods such as multivariate microaggregation provide low usability due to the complexity introduced by its NP-hard nature~(\cite{torra2017fuzzy}). Data condensation provides an efficient solution to privacy preservation of data streams; however, the quality of data degrades as the data grows, eventually leading to low utility~(\cite{bertino2005framework}). Many of the multi-dimensional approaches, such as rotation perturbation and geometric perturbation, introduce high computational complexity and take unacceptably long time to execute~(\cite{chen2005random,chen2011geometric}). This means that such methods in their default settings are not feasible for high dimensional data such as big data and data streams. 
A structured approach is needed, which can provide a practically applicable solution for selecting the best privacy preservation approach for a given application or dataset.

Several works have looked at the connection between privacy, utility, and usability.  Bertino et al. proposed a framework for evaluating privacy-preserving data mining algorithms; for each algorithm, they focused on assessing the quality of the sanitized data ~(\cite{bertino2005framework}).  Other frameworks aim at providing environments for dealing with sensitive data. Sharemind is a shared multi-party computation environment allowing secret data-sharing ~(\cite{bogdanov2008sharemind}). FRAPP is a matrix-theoretic framework aimed at helping the design of privacy-preserving random perturbation schemes  ~(\cite{agrawal2005framework}).  Thuraisingham et al. went one step further; they provide a vision for designing a framework that measures both the privacy and utility of multiple privacy-preserving techniques. They also provide insight into balancing privacy and utility in order to provide better privacy preservation ~(\cite{thuraisingham2017towards}). However, these frameworks neither provide a solution to the problem of dealing with numerous privacy preservation algorithms  nor provide proper quantification of data utility and privacy for a particular application of the dataset at hand.

\section{Background}
Choosing the most appropriate data perturbation algorithm out of many algorithms is the primary challenge addressed by the proposed approach. The proposed framework named PPaaS aims to select the optimal perturbation algorithm that, when applied to a particular dataset, provides a proper balance between privacy and utility. This section discusses the different components involved in the conceptual development of PPaaS. Data perturbation is the process of modifying data using a certain mechanism (e.g. noise addition, geometric transformation, randomization) to prevent third parties from identifying the owners of data while performing important data analytics. A perturbed dataset's key property is its indistinguishability from the original data due to maintaining the same format. A third-party would not have the immediate impression of accessing a different dataset than the original dataset, as opposed to accessing an encrypted database. However, as all analytics are carried out on perturbed data, a certain utility level of the perturbed data must be maintained. Hence, enabling enough utility while maintaining enough privacy is the main challenge in data perturbation, in other words,  proper balance between privacy and utility has to be maintained. Due to these dynamics, a perturbed dataset is expected to leak some information. A privacy model theoretically defines the level of privacy offered by a perturbation algorithm. It is important to understand the theoretical guarantees provided by a particular perturbation algorithm, as it provides an initial impression of the robustness of the corresponding perturbation algorithm. Besides, perturbed datasets can be vulnerable to data reconstruction attacks (a type of privacy attack). A data reconstruction attack tries to exploit the information leaked from a perturbed dataset with the purpose of reconstructing the original data. It is also essential to identify the resistance of a specific perturbed dataset to data reconstruction attacks, in order to provide an empirical guarantee towards the robustness of the perturbation applied to the dataset.

\subsection{Perturbation Algorithms}

\begin{figure}[H]
\centering
\scalebox{0.4}{
\includegraphics[width=1\textwidth, trim=0cm 0cm 0cm 0cm]{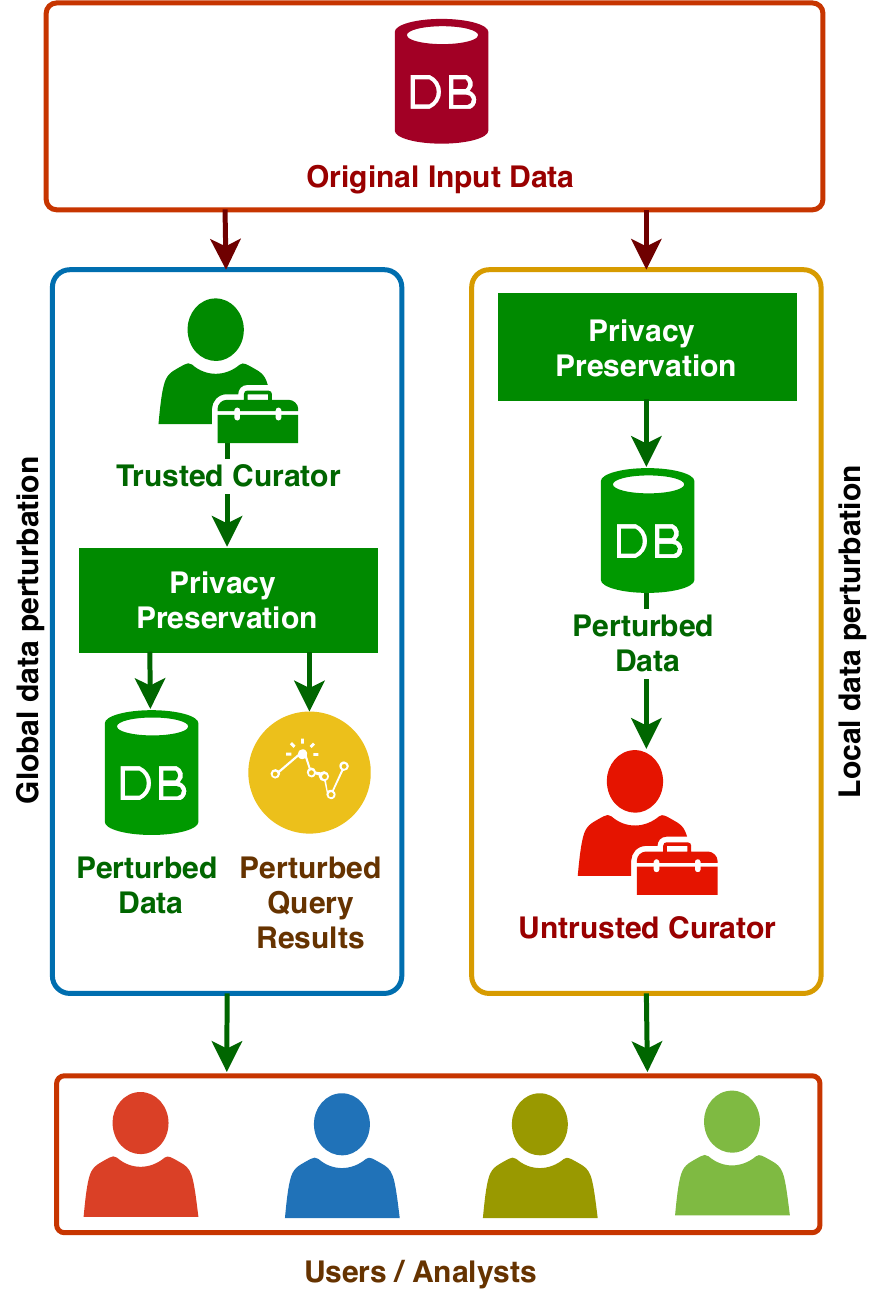}
	}
\caption[Global (output) Vs. local (input) data perturbation]{Global data perturbation Vs. Local data perturbation. 
}
\label{locglobdiff}
\end{figure}

A data perturbation algorithm's main functionality is to modify the original input data before releasing them to a third-party (e.g. an analyst) to limit possible privacy leaks or attacks by adversaries. Since the modified data are not subjected to any format conversion as in data encryption, data perturbation has lower time and space complexity compared to cryptographic approaches (e.g. fully homomorphic encryption), which are used to enforce a high level of privacy. However, for the same reason, data perturbation still results in a certain level of data leak, which needs to be carefully evaluated to restrict unanticipated privacy leaks. Data perturbation can be categorized into two classes: (1) input data perturbation, and (2) output data perturbation. Input data perturbation is also called local data perturbation, whereas output data perturbation is also called global data perturbation. As shown in Figure \ref{locglobdiff}, in input data perturbation approaches (represented on the right-hand side of the figure), data perturbation is performed on the data when they leave the data owners.

In output data perturbation approaches (represented on the left-hand side of the figure), a trusted curator applies data perturbation to analysis query results that were obtained by running queries on original data. Both input and output perturbation are often used. However, in an untrusted setting where only malicious parties are present, input data perturbation is preferred. Input perturbation applies higher randomization, hence input perturbation is considered to provide better privacy than output perturbation~\cite{kairouz2014extremal}, while, for the same reason, output perturbation often produces better utility. Input perturbation can be divided further into unidimensional perturbation and multidimensional perturbation ~\cite{okkalioglu2015survey}. Additive perturbation ~\cite{muralidhar1999general}, microaggregation~\cite{torra2017fuzzy}, randomized response ~\cite{dwork2014algorithmic}, rounding~\cite{hundepool2012statistical}, data swapping ~\cite{hasan2016effective} and resampling~\cite{martinez2012towards}  are examples of unidimensional input perturbation. Data condensation~\cite{aggarwal2004condensation}, random rotation ~\cite{chen2005random}, geometric perturbation~\cite{chen2011geometric}, random projection ~\cite{liu2006random},  sketch-based approach~\cite{aggarwal2007privacy} are a few examples of multidimensional perturbation approaches. The merge of several forms of perturbation together is referred to as hybrid perturbation~\cite{aldeen2015comprehensive}.  Output perturbation is achieved using approaches such as noise addition~\cite{abadi2016deep} and exponential mechanism~\cite{jones2019towards}. In PPaaS, we use the trusted curator scenario as PPaaS wants a trusted curator to conduct the process of selecting the best perturbation instance out of multiple perturbation instances based on the dynamics of the original dataset. However, PPaaS employs input perturbation algorithms in applying perturbation over input datasets, where the perturbation is conducted over the entire dataset at once.

\subsection{Privacy Models}

A privacy model/privacy definition specifies the limits of private information disclosure by a certain perturbation mechanism~\cite{machanavajjhala2015designing}; $k-anonymity$, $l-diversity$, $(\alpha, k)-anonymity$, $t-closeness$~\cite{chamikara2019efficient2,li2007t} are examples of earlier privacy models. A database provides $k-anonymity$ if the data is indistinct from a minimum of $(k-1)$ records~\cite{sweeney2002k}. $l-diversity$ was introduced to overcome the issues of $k-anonymity$ by improving the diversity and to reduce the homogeneity of sensitive attributes~\cite{machanavajjhala2007diversity}. A $k-anonymous$ database provides $l-diversity$ if each equivalent class has at least $l$ well-represented values for each sensitive attribute~\cite{machanavajjhala2007diversity}. A database provides $t-closeness$ if the difference of values between a sensitive attribute in any equivalence class and the distribution of the attributes in the whole database is no more than $t$~\cite{li2007t}. However,  these models and their improvements show vulnerability to privacy attacks such as minimality attack ~\cite{zhang2007information}, composition based attacks~\cite{ganta2008composition}, and foreground knowledge~\cite{wong2011can}. Compared to previous privacy definitions, differential privacy (DP) provides a strong privacy model that is trusted to provide a better level of privacy guarantee compared to previous privacy models ~\cite{dwork2009differential, mohammed2011differentially,fan2020privacy,wang2020deep}. A randomized algorithm $\mathcal{A}$  satisfies $\varepsilon$-LDP if for all pairs of users' inputs $v_1$ and $v_2$ and for all $\mathcal{Q} \subseteq Range(\mathcal{A})$, and for ($\varepsilon \geq 0$)  Equation \eqref{ldpeq} holds. $Range(\mathcal{A})$ is the set of all possible outputs of the randomized algorithm $\mathcal{A}$.

\begin{equation}
\mathcal{P}r[\mathcal{A}(v_1) \in \mathcal{Q}] \leq \exp(\varepsilon)~Pr[\mathcal{A}(v_2) \in \mathcal{Q}]
\label{ldpeq}
\end{equation}

\subsection{Privacy Attacks}
Effective noise reconstruction techniques can significantly reduce the level of privacy in additive perturbation techniques ~\cite{okkalioglu2015survey}.  Data perturbation approaches are vulnerable to various data reconstruction attacks such as naive estimation, independent component analysis (ICA)-based attacks, known I/O attacks, eigenanalysis, distribution analysis attacks, and spectral filtering~\cite{chen2007towards,liu2008survey}. A data reconstruction attack tries to reconstruct original input data from the perturbed data. Naive estimation explores the difference between perturbed and original data. Hence, a strong perturbation can provide enough resistance to naive inference. ICA-based attacks employ independent component analysis to reconstruct the original data~\cite{chen2007towards}. Known I/O attacks assume that the attacker knows/has a specific portion of the original data and knows the mapping between the known data and its corresponding perturbed data~\cite{chen2007towards}. The attacker can try to use this knowledge of mapping to reconstruct the rest of the original data from the perturbed data~\cite{chen2007towards}. Eigenanalysis-based attacks try to filter out the random noise from the perturbed data by analyzing the eigenvectors of the data. Spectral filtering, singular value decomposition (SVD) filtering, and principal component analysis (PCA) filtering are three examples of eigenanalysis-based attacks~\cite{agrawal2000privacy}. Distribution analysis attacks try to reconstruct the probability density function of the original data~\cite{chen2007towards}. For example, microaggregation to a single variable (univariate microaggregation) is vulnerable to transparency attacks when the published data includes information about the protection method and its parameters ~\cite{torra2017fuzzy}.  These attacks demand that new data perturbation approaches be more robust in advanced adversarial settings to provide sufficient resilience against such attacks.

\subsection{Fuzzy Inference Systems}
\label{fissection}
\begin{figure}[H]
	\centering
	\scalebox{0.7}{
	\includegraphics[width=1.1\textwidth, trim=0.5cm 0cm 0.5cm
	 0cm]{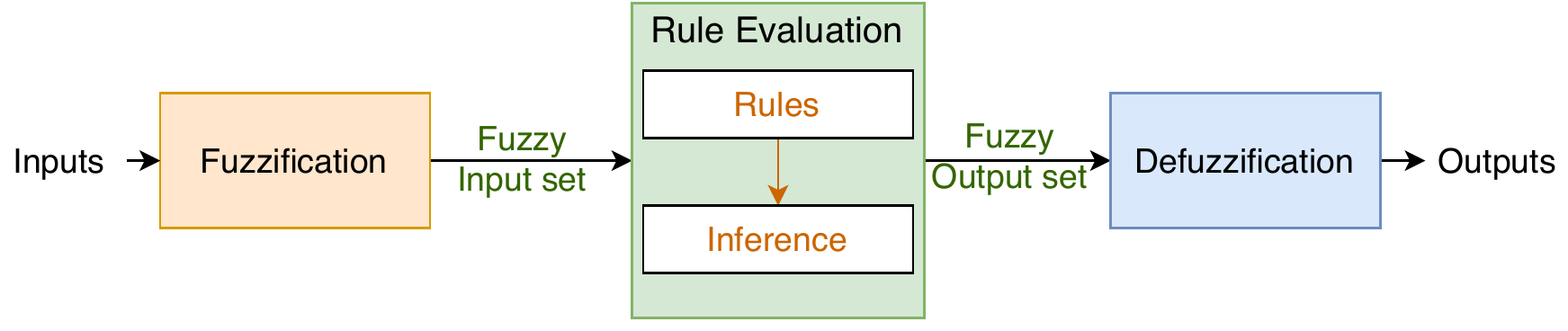}
	 }
	\caption{Basic flow of a fuzzy inference system}
	\label{fis}
\end{figure}

The proposed framework (PPaaS) uses fuzzy logic to derive a final score (named as the fuzzy index: FI) to the overall quality of privacy, attack resistance, and utility of a perturbed dataset. Fuzzy logic is a logical system that provides the capability to design real-world problems as human-thinking-oriented computational paradigms. Fuzzy logic is a precise approach that allows the modeling of impreciseness of the features such as big, hot, and slow, using a multi-valued approach as opposed to classical logic that is based on 1 and 0 (binary).  Hence, fuzzy logic is termed as a precise logic of imprecision. As shown in Figure \ref{fis},  fuzzification, rule evaluation, and defuzzification are the three main steps of a conventional fuzzy inference system. 
Fuzzy logic has been used to solve many real-world problems, including fuzzy automatic transmission, hand-writing recognition, and voice recognition. Besides, fuzzy logic can be used as an effective tool for ranking-based algorithms~\cite{gupta2015new,tran2008qos}. PPaaS utilizes this capability of fuzzy logic to generate ranks for the overall quality of privacy, attack resistance, and utility of a perturbed data instance. 

\section{Privacy Preservation as a Service}

We propose a novel approach named ``Privacy Preservation as a Service (PPaaS)", a generic framework that can be used to sanitize big data in a granular and application-specific manner. In this section, we give a detailed outline of the concept. The high diversity and specificity of privacy preservation methods presents complexities, such as finding a trade-off between privacy, attack resistance, and utility. As noted in Section \ref{litrev}, privacy preservation algorithms can suffer from different types of biases. For example, a particular sanitization algorithm used for privacy-preserving classification may not have DM bias, but it may suffer from Type B and D biases, while another one has only Type B bias, and a third one has DM bias. 
Different applications may tolerate different types of bias, and there is no general rule. These biases govern the utility of a perturbed dataset. Besides, a perturbed dataset can be vulnerable to different data reconstruction attacks. It is essential to investigate the attack resistance of a perturbed dataset. A mechanism that identifies the best perturbed instance of an input dataset based on privacy, attack resistance, and utility is essential.

\begin{figure}[H]
	\centering
	\scalebox{0.7}{
	\includegraphics[width=1.1\textwidth, trim=0.5cm 0cm 0.5cm
	 0cm]{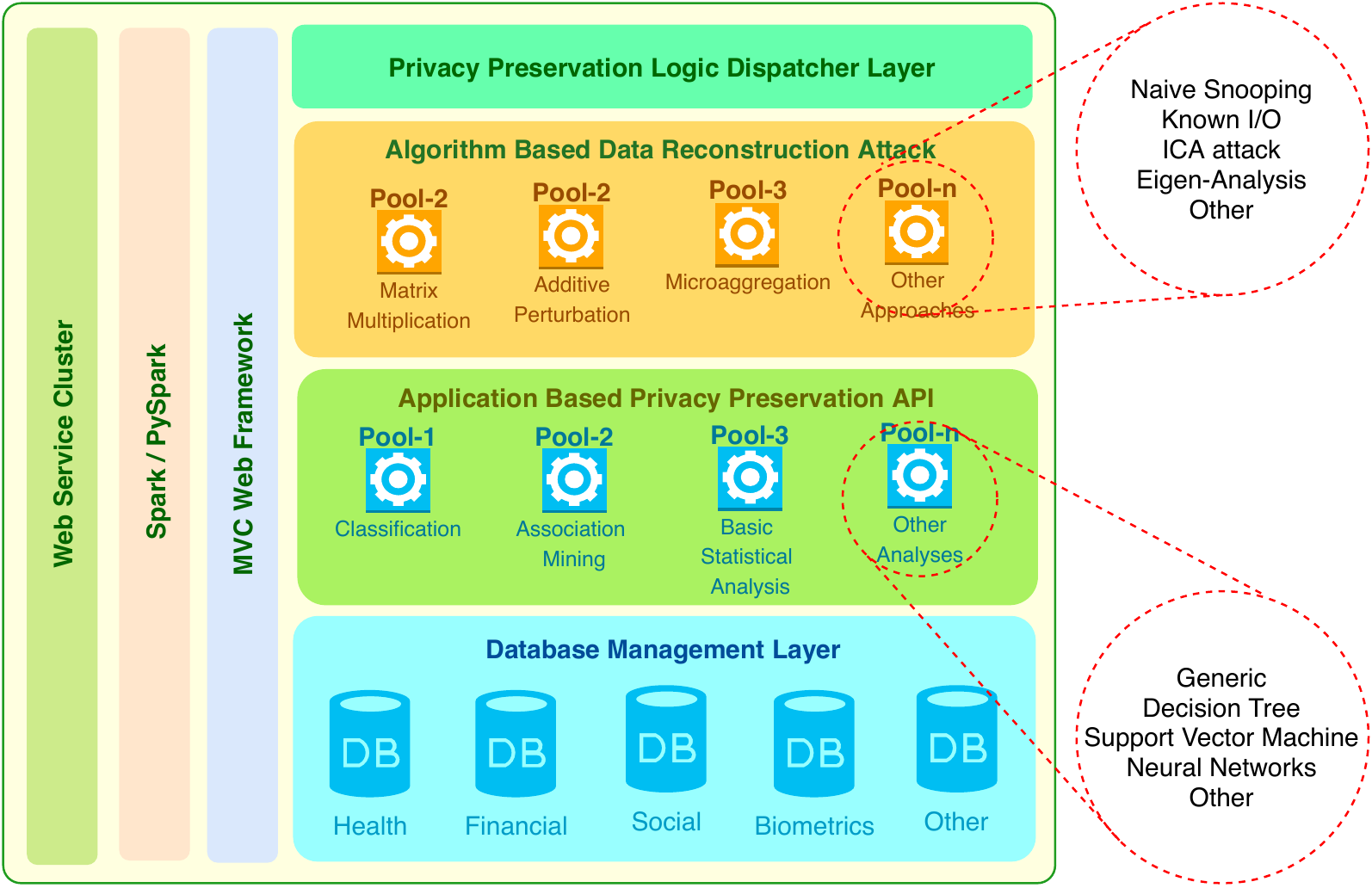}
	 }
	\caption{Privacy preservation as a service (PPaaS) for big data.}
	\label{framework}
\end{figure}

The primary intuition of PPaaS is that different privacy preservation algorithms are suitable for different data owner requirements (privacy and performance) and different data requester needs (utility and usability). A unified service of data sanitization for big data can provide an interactive solution for this problem.
PPaaS can choose the most suitable privacy preservation algorithm for a particular analysis at hand. The architecture of PPaaS is presented in Figure \ref{framework}. It is implemented as a web-based framework that can operate in a web service cluster. The scalability necessary for big data processing is achieved using APIs such as Spark/PySpark~(\cite{drabas2017learning}) (as the primary implementation language was Python) with a clean build design adapted with a Model-View-Controller (MVC) web framework. As Figure \ref{framework} shows, the framework consists of four distinct layers: (1) the database management layer, (2) the application-based privacy preservation API, (3) the algorithm-based data reconstruction attack layer, and (4) the privacy preservation logic dispatcher layer.

\begin{figure}[H]
	\centering
	\scalebox{0.7}{
	\includegraphics[width=1.1\textwidth, trim=0.5cm 0cm 0.5cm
	 0cm]{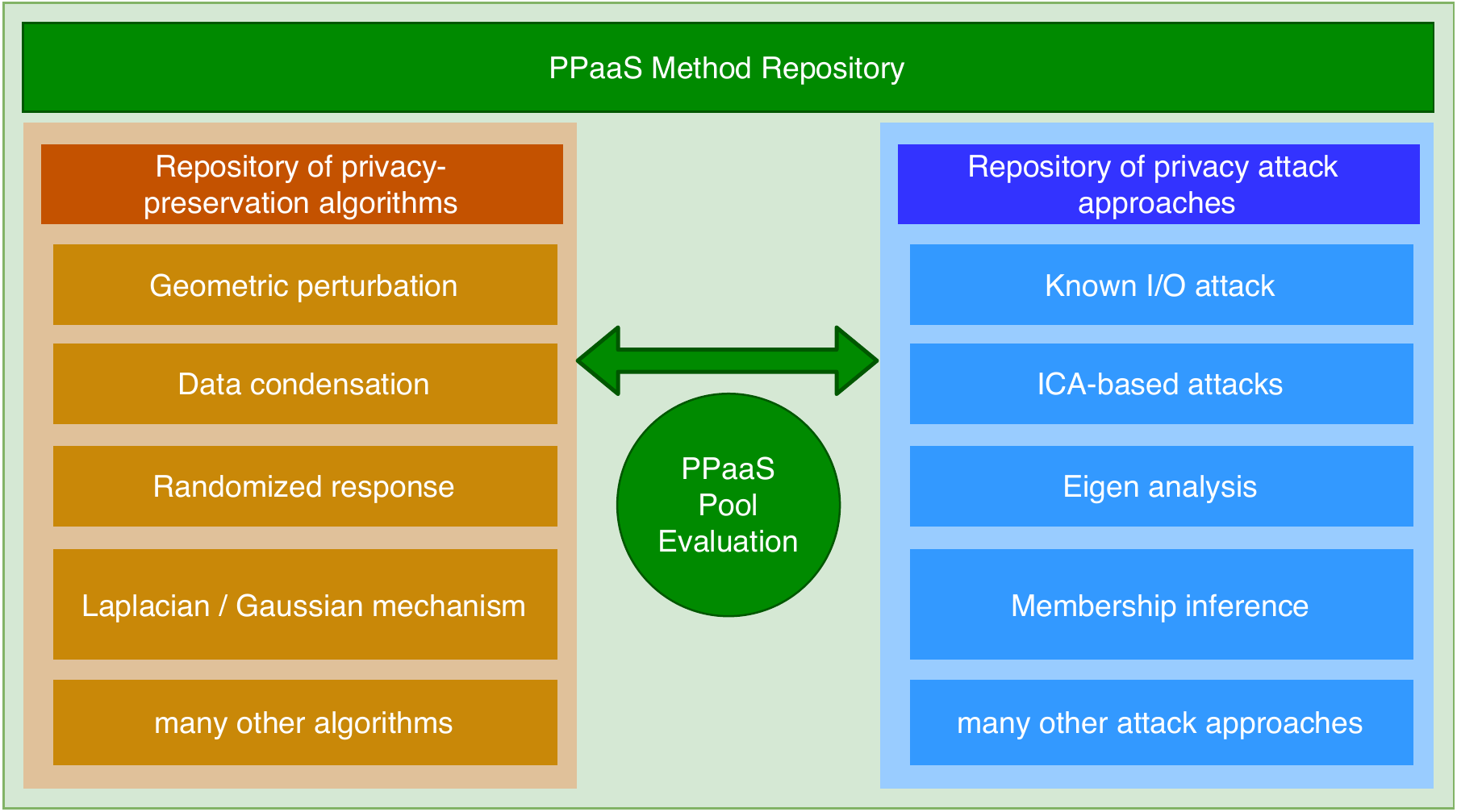}
	 }
	\caption{PPaaS Method Repository}
	\label{ppaasrepository}
\end{figure}

The privacy preservation module consists of pools of application logic (e.g. classification and association mining), and pools of privacy preservation algorithms (e.g. matrix multiplication, additive perturbation) and pools of data reconstruction attacks.  The PPaaS privacy preservation module integrates a collection of privacy preservation algorithms into a collection of pools where each pool represents a particular class of data mining/analysis algorithms. The enlargement of the red circle in Figure \ref{framework} shows a possible collection of sub-pools of privacy preservation algorithms for classification.  For instance, rotation perturbation (RP)~(\cite{chen2005privacy}) can be integrated into the "Generic" sub-pool of pool1: Classification (refer to the red circles in Figure \ref{framework}), as it provides better accuracy towards a collection of classification algorithms. A particular pool may have several subdivisions to enable the synthesis of new data sanitization methods that are tailored to more specific requirements. Similarly, the pools of data reconstruction attacks are also listed to be tested against perturbed data instances of input datasets. The database management layer provides the necessary services for uniform data formatting. Figure \ref{ppaasrepository} represents the pool engagements between privacy preservation algorithms (data perturbation) and data reconstruction attack approaches. During the analysis, PPaaS selects the corresponding data perturbation approaches and the related data reconstruction attacks based on the corresponding pool classifications. In the proposed concept, privacy preservation is discussed in terms of data perturbation. The following sections use "privacy preservation" and "perturbation" interchangeably, referring to the same objective.

A data owner/curator  can utilize the framework to impose privacy on a particular dataset for a particular application by using the best privacy preservation approach from a pool of available algorithms. In the proposed setting, PPaaS requires a trusted curator to identify the query or the analysis requests for a given dataset, and run the PPaaS logic for the corresponding application (e.g. deep learning~(\cite{lecun2015deep})). 
The curator/data owner accesses the data and applies privacy preservation (perturbation) to the data or  dataset according to the users' requirements.

\begin{figure}[H]
	\centering
	\scalebox{1}{
	\includegraphics[width=0.7\textwidth, trim=0.5cm 0cm 0.5cm
	 0cm]{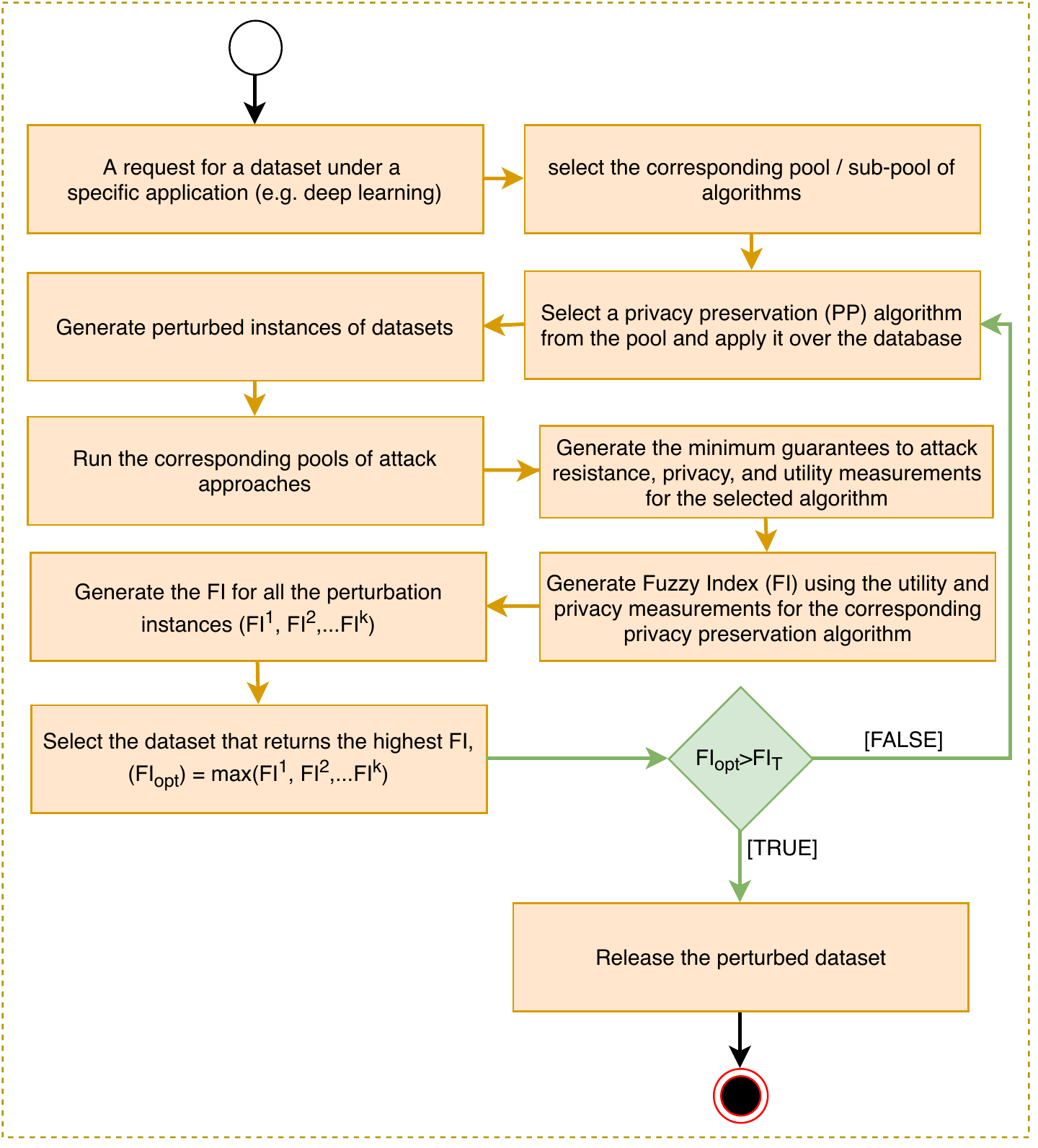}
	 }
	\caption{Flow of events in application specific privacy preservation of PPaaS}
	\label{mainflow}
\end{figure}

PPaaS investigates three key aspects: (1) understanding the data owner/producer requirements for privacy, (2) understanding the data requester/consumer utility needs, and  (3) selecting and applying the optimum privacy-preserving algorithm to the data. Next, the perturbed instances of the input dataset are tested against the most suitable pool of data reconstruction/privacy attacks. Besed on the performance, the attack experiment analysis generates a minimum guarantee of resistance(MGR). Finally, the result of applying privacy preservation to a particular dataset is assessed using a FIS (Fuzzy Inference System) based fuzzy metric (named the fuzzy index or FI), which is a single metric to evaluate the balance between privacy and utility provided by the corresponding privacy preservation algorithm.  Fig \ref{mainflow} shows the main flow of PPaaS in releasing a perturbed dataset with a customized application of privacy-preservation. The data curator will receive a request for a certain operation on the underlying dataset. For example, this request can be for deep learning on a medical dataset that is maintained by the corresponding data owner. 
The data owner forwards the request to the PPaaS framework, which will select the corresponding pool/sub-pool of privacy preservation algorithms allocated under deep learning. In the example, this pool may include the following algorithms: local differentially private approaches, geometric data perturbation approaches, random projection-based data perturbation approaches, which are suitable for producing high utility for deep learning. Next, PPaaS sequentially applies the corresponding pool of privacy preservation algorithms. Then, PPaaS runs the corresponding pool of data reconstruction algorithms on each of the perturbed data instances to generate a minimum guarantee of attack resistance for each perturbed dataset. Based on the results, PPaaS generates a fuzzy index for each perturbed data instance (perturbation algorithm).  If a particular pool has four privacy preservation algorithms, PPaaS will produce for perturbed data instances, which will result in 4 FI values.  Next, the PPaaS will select the perturbed dataset with the highest FI, because the corresponding dataset provides the best balance between privacy, attack resistance, and utility.

PPaaS uses a fuzzy inference system (FIS) to generate the fuzzy index. Privacy, minimum guarantee of attack resistance (MGR) and utility are the only inputs to the FIS that generates a final score, that is, the fuzzy index ($FI$). $FI$ is a quantitative rank that rates the complete process of privacy preservation upon a particular dataset for a given application. A heuristic approach was followed in defining the fuzzy rules that focused on maintaining a balance between privacy, attack resistance, and utility. The universe of discourse of the inputs and output ranges from 0 to 1.  A higher FI value suggests that the final dataset has high privacy, attack resistance, and utility with a good balance between them. The PPaaS dispatcher investigates the value of $FI$ corresponding to a particular process of sanitization, compares it with a user-defined balance guarantee, $ FI_T$ that is taken as an input parameter from the data owner. If $FI_{opt}\geqslant FI_T$, the dataset will be released to the data requester, where $FI_{opt}$ is the maximum $FI$ generated by the pool. Otherwise, the PPaaS will reapply the random perturbation algorithm to find a better solution that satisfies $FI_T$ requirement.

\begin{figure}[H]
	\centering
	\scalebox{0.6}{
	\includegraphics[width=1.1\textwidth, trim=0.5cm 0cm 0.5cm
	 0cm]{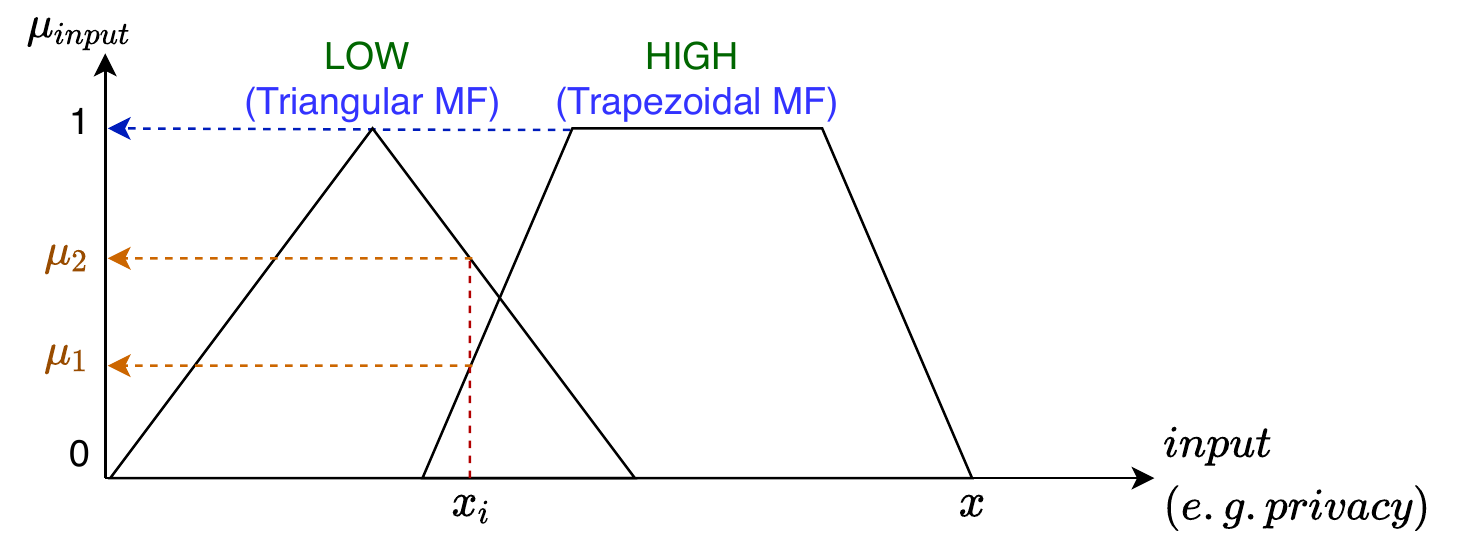}
	 }
	\caption{An instance of an input variable definition}
	\label{inputvariable}
\end{figure}

A fuzzy inference system (FIS) takes several inputs and generates a certain output based on evaluating a collection of specified rules, which are expressed as fuzzy rules (refer to Section \ref{fissection}). In an FIS, the first step is to apply fuzzification to the input variables. Fuzzification maps inputs to values from 0 to 1 using a collection of membership functions. There are different types of membership functions that can be used for this step. Triangular, Trapezoidal, Piecewise linear, Gaussian, and Singleton are some examples of such membership functions. The most suitable membership function and its range and shape for a particular problem need to be selected based on the problem's dynamics in the corresponding environment and the domain expert's knowledge.  Figure \ref{inputvariable} represents the fuzzification of an input variable (e.g. privacy) using two membership functions that represent two levels (LOW and HIGH) of the input.  LOW is represented using a triangular membership function, whereas HIGH is represented using a Trapezoidal MF. In this plot (refer to Figure \ref{inputvariable}), $\mu_{input}$ quantifies the corresponding input's ($x_i$) degree of membership. 

\begin{figure}[H]
	\centering
	\scalebox{0.35}{
	\includegraphics[width=1\textwidth, trim=0.5cm 0cm 0.5cm
	 0cm]{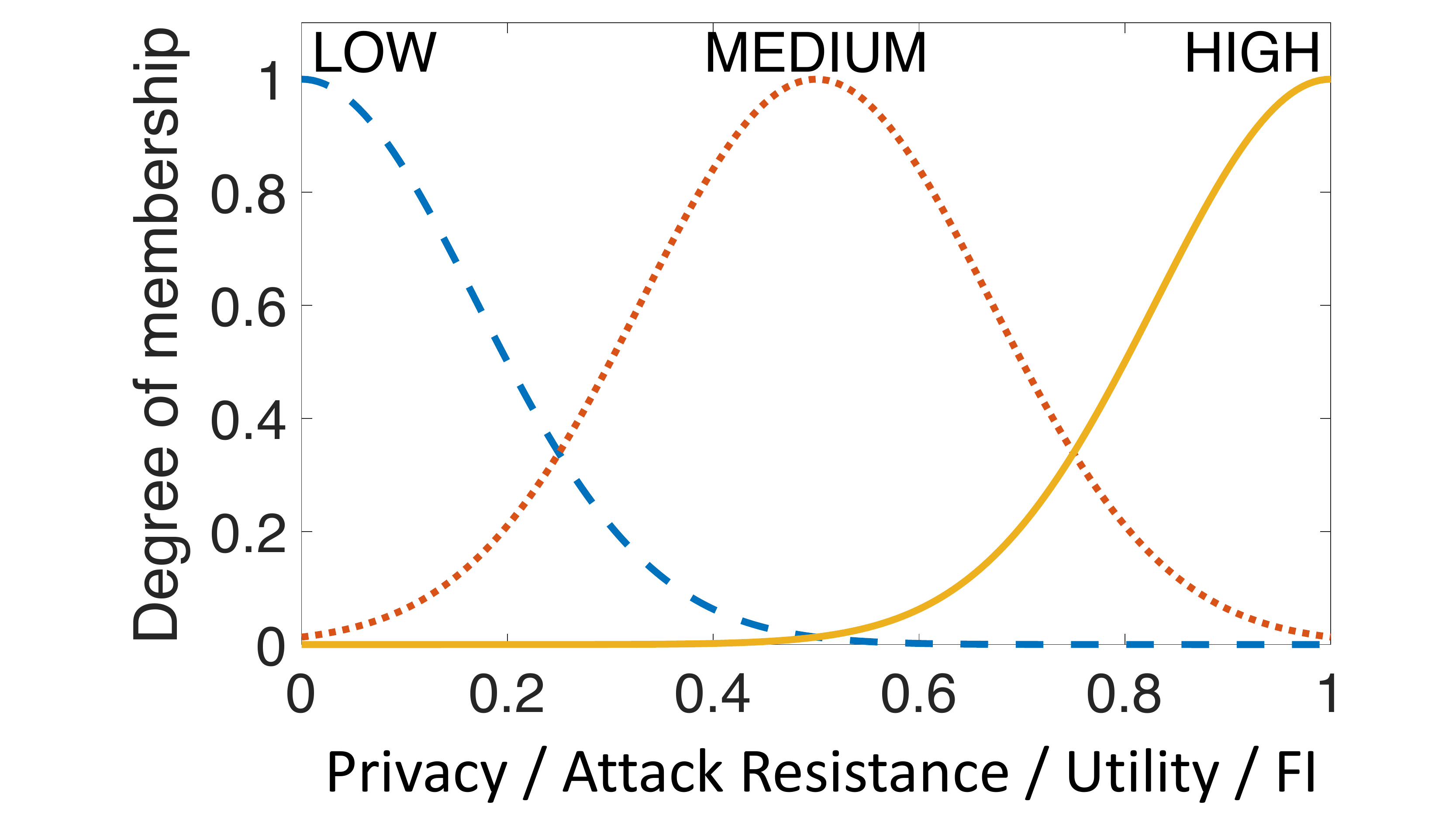}
	 }
	\caption{Fuzzy membership functions of the input/output variables}
	\label{fuzzyfunc}
\end{figure}

We need to first map privacy, attack resistance, and utility into fuzzy memberships. Figure \ref{inputvariable} shows an example of mapping input values into fuzzy memberships, which allows the activation of rules that are in terms of linguistic variables. During this process, the membership functions allow the fuzzifier to determine the degree to which the input values belong to each membership function. For example, the figure shows the fuzzification of the input value, $x_i$, which produces the two fuzzified membership (degree of membership) values , LOW($x_i$)=$\mu_1$ and HIGH($x_i$)=$\mu_2$. Figure \ref{fuzzyfunc} shows the mapping of the three inputs: privacy, attack resistance, utility, and the output: FI into fuzzy memberships (A more detailed explanation on PPaaS input and output fuzzification is included later). 

\begin{figure}[H]
	\centering
	\scalebox{0.2}{
	\includegraphics[width=1.1\textwidth, trim=0.5cm 0cm 0.5cm
	 0cm]{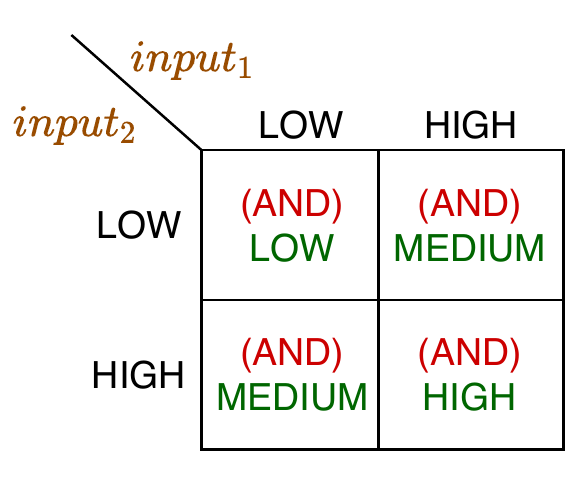}
	 }
	\caption{Example rule base of a fuzzy inference system}
	\label{rulebase}
\end{figure}

As shown in Figure \ref{fis}, the second module of an FIS is the rule evaluation, which involves inference based on a collection of linguistic rules, which is also called the rule base.  A rule is defined using the IF-THEN convention (e.g. IF $input_1$ is HIGH AND $input_2$ is LOW THEN output is MEDIUM ). As shown in Figure \ref{fuzzyfunc}, we can identify that all inputs and outputs of PPaaS has three levels of memberships (LOW, MEDIUM, HIGH). Assume that a particular FIS has two inputs ($input_1$ and $input_2$) and one output. If $input_1$ and $input _2$ have two membership levels each (e.g. LOW and HIGH) and the output with the levels LOW, MEDIUM, and HIGH, we can define an example rule-base, which is shown in Figure \ref{rulebase}. Each box in the figure represents a rule where the value of a box represents the output variable's membership for the corresponding rule.  

\begin{figure}[H]
	\centering
	\scalebox{0.5}{
	\includegraphics[width=1.1\textwidth, trim=0.5cm 0cm 0.5cm
	 0cm]{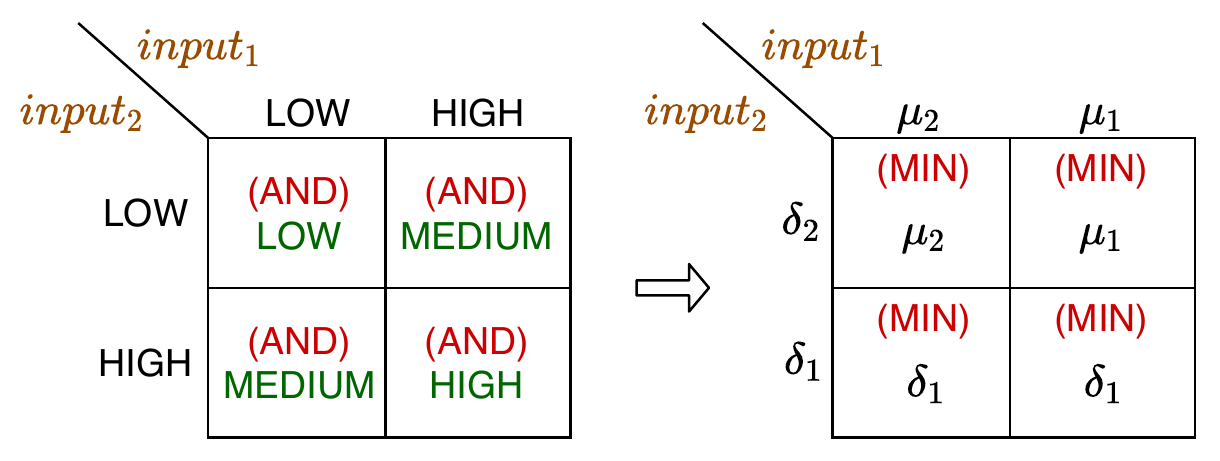}
	 }
	\caption{Rule evaluation based on the rule base}
	\label{ruleevaluation}
\end{figure}

The rule evaluation step (the inference engine) combines all fuzzy conclusions obtained by inferencing the rules into a single conclusion. Each inference will suggest a different action. A simple MAX-MIN (OR-AND) operation of the selection can be used where the maximum fuzzy value of the inference is generally used as the final conclusion. For simplicity, let us consider both $input_1$ and $input_2$ have the same function definitions represented in Figure \ref{inputvariable}.  Let us consider that $\delta_1$, $\delta_2$ are the two membership values returned by LOW and HIGH membership functions of $input_2$ and $\delta_1<\mu_1<\mu_2<\delta_2$. Assuming that we consider the AND (equivalent to MIN) operation between antecedents, we can obtain the final values for the output membership functions as represented in Figure \ref{ruleevaluation}. For the rules with the same consequent, the OR (MAX) operation between the corresponding consequent values is generally considered. 

\begin{figure}[H]
	\centering
	\scalebox{0.7}{
	\includegraphics[width=1.1\textwidth, trim=0.5cm 0cm 0.5cm
	 0cm]{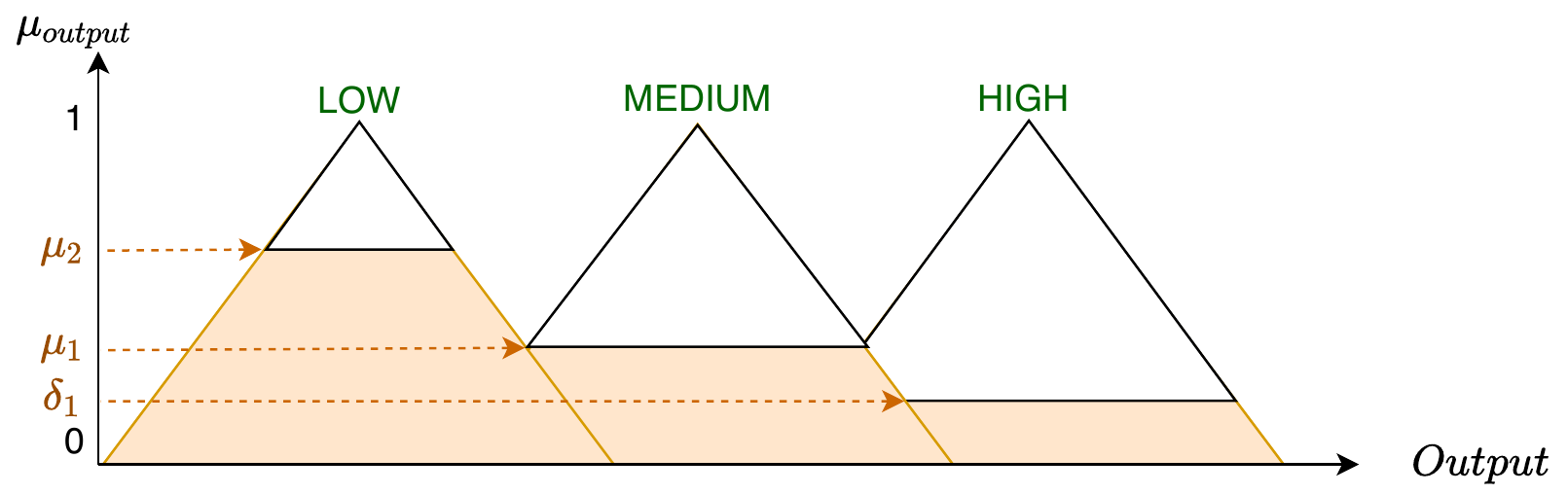}
	 }
	\caption{Rule aggregation on the output variable}
	\label{outputvariable}
\end{figure}

As shown in Figure \ref{outputvariable}, we can use clipping ($\alpha$-cut) and aggregate the rules to obtain the colored area in the membership levels in the output variable. In this figure, $\mu_{output}$ represents the output membership values ($\mu_1$, $\mu_2$, and $\delta_1$) obtained by the rule evaluation process (as depicted by Figure \ref{ruleevaluation}).  The final step of the fuzzy inference system is to apply defuzzification based on the aggregated shape of the output function. There are several defuzzification approaches; however, the most popular approach is the centroid-based technique, which finds the point where a vertical line would slice the aggregate set into two equal masses (the center of gravity: COG).  Equation \ref{cogequation} can be used to obtain this value (COG), where $x = output$ and $\mu_{x} = \mu_{output}$ (refer to Figure~\ref{outputvariable}).

\begin{equation}
COG=\frac{\int_{min}^{max}  \mu_{x} x dx}{\int_{min}^{max} \mu_{x} dx}
\label{cogequation}
\end{equation}

\begin{figure}[H]
	\centering
	\scalebox{1}{
	\includegraphics[width=1 \textwidth, trim=0.5cm 0cm 0.5cm
	 0cm]{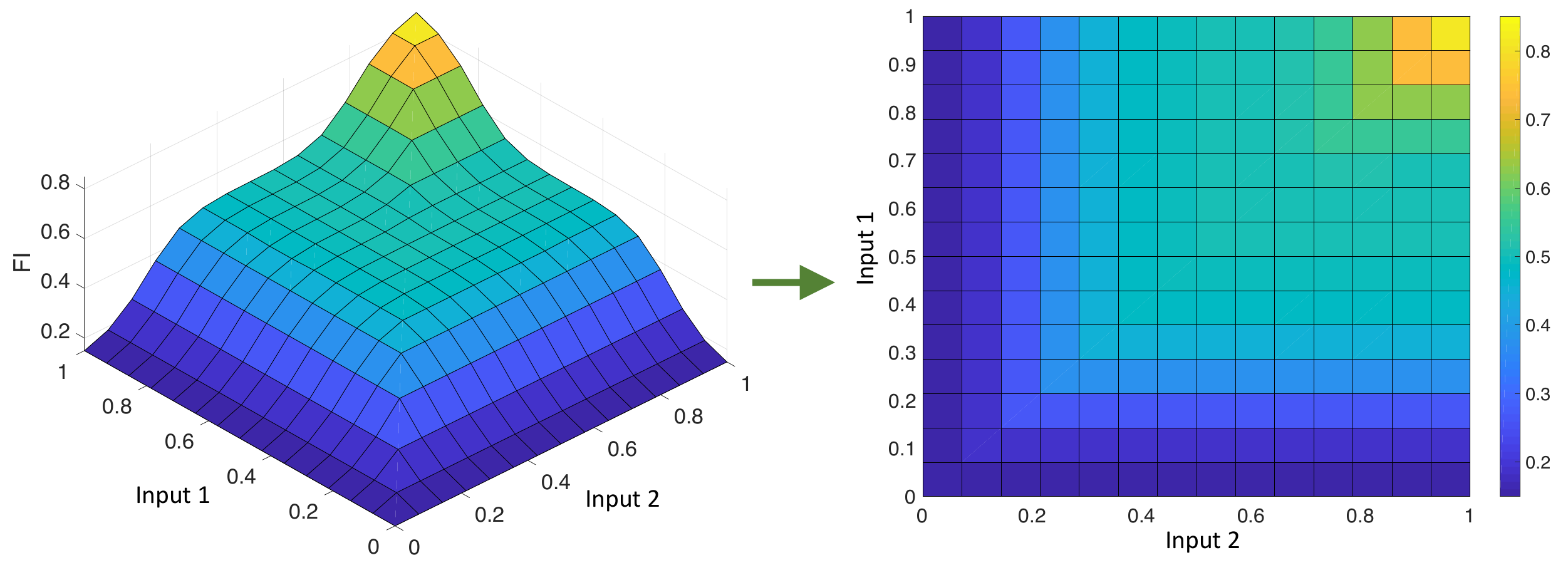}
	 }
	\caption{Rule surface of the FIS}
	\label{rulesurf}
\end{figure}

In the proposed framework (PPaaS), we define a FIS to take the three inputs: the minimum guarantee of privacy, MGR, and the minimum guarantee of utility to produce an output named fuzzy index ($FI$). $FI$ provides an impression of the quality of the balance between privacy, attack resistance, and utility generated after perturbing a dataset using a privacy preservation algorithm. According to the domain knowledge, we already know that a good privacy preservation algorithm should enforce high privacy, high attack resistance while producing good utility (e.g. accuracy). Following this notion, $FI$ should ideally provide high values only when the minimum guarantee of privacy, MGR, and the minimum guarantee of utility are high. In case one is high and the other is low, the $FI$ should be a lower value. Hence, the fuzzy model should produce a rule-surface, as presented in Figure \ref{rulesurf}. Considering all these dynamics between privacy, utility, and $FI$, we introduced three membership functions (LOW, MEDIUM, HIGH) for each variable. Next, we considered Gaussian functions for all the membership functions in the two input variables and output variables, as shown in Figure \ref{fuzzyfunc}. Finally, we defined the eleven rules given in Equation \ref{rules} to obtain the rule-surface depicted in Figure \ref{rulesurf}. As defined in the first three rules, the fuzzy inference engine will generate a low value for $FI$ when any one of the three input parameters (``privacy", ``attack\_resistance", and ``utility") take a low input value. This is to ensure that the lower the value of any one of the input parameters, the lower the $FI$ value. Consequently, the remaining rules do not consider any rule combination where any one of the input parameters is LOW.  Rule 5 to Rule 11 consider the dynamics of $FI$ under all remaining combinations of membership levels ($MEDIUM$ and $HIGH$). For example, Rule 4 considers the situation where ``privacy", ``attack\_resistance", and ``utility" are $MEDIUM$, $MEDIUM$, and $MEDIUM$ respectively, and under this situation $FI$ is considered to result in $MEDIUM$ values.

\begin{equation}
\scalebox{0.7}{$
\begin{aligned}
& \textbf{Rule 1:}\ IF (privacy = LOW) \ THEN\ (FI = LOW) \\
& \textbf{Rule 2:}\ IF (attack\_ resistance = LOW) \ THEN\ (FI = LOW) \\
& \textbf{Rule 3:}\ IF (utility = LOW) \ THEN\ (FI = LOW) \\
& \textbf{Rule 4:}\ IF (privacy = MEDIUM\ AND\ attack\_ resistance = MEDIUM\ AND \ utility = MEDIUM) \ THEN\ (FI = MEDIUM) \\
& \textbf{Rule 5:}\ IF (privacy = MEDIUM\ AND\ attack\_ resistance = MEDIUM\ AND \ utility = HIGH) \ THEN\ (FI = MEDIUM) \\
& \textbf{Rule 6:}\ IF (privacy = MEDIUM\ AND\ attack\_ resistance = HIGH\ AND \ utility = MEDIUM) \ THEN\ (FI = MEDIUM) \\
& \textbf{Rule 7:}\ IF (privacy = MEDIUM\ AND\ attack\_ resistance = HIGH\ AND \ utility = HIGH) \ THEN\ (FI = HIGH) \\
& \textbf{Rule 8:}\ IF (privacy = HIGH\ AND\ attack\_ resistance = MEDIUM\ AND \ utility = MEDIUM) \ THEN\ (FI = MEDIUM) \\
& \textbf{Rule 9:}\ IF (privacy = HIGH\ AND\ attack\_ resistance = MEDIUM\ AND \ utility = HIGH) \ THEN\ (FI = HIGH) \\
& \textbf{Rule 10:}\ IF (privacy = HIGH\ AND\ attack\_ resistance = HIGH\ AND \ utility = MEDIUM) \ THEN\ (FI = HIGH) \\
& \textbf{Rule 11:}\ IF (privacy = HIGH\ AND\ attack\_ resistance = HIGH\ AND \ utility = HIGH) \ THEN\ (FI = HIGH) 
\end{aligned}
$}
\label{rules}
\end{equation}

Figure \ref{rulesurf} depicts the rule surface of the fuzzy inference system (FIS), which is used to generate $FI$. It shows the change of $FI$ when any two of the inputs (``privacy", ``attack\_resistance", and ``utility") are varied while the third input is kept constant. Consequently, when Input 1 is ``privacy", Input 2 can be ``attack\_resistance" or ``utility".  As shown in the figure, FIS generates higher values for $FI$ when both utility and privacy are high, whereas for lower values of privacy and utility $FI$ also stays at a lower level. As shown in the figure, the rule surface makes sure that a higher value of only one parameter (privacy, MGR, or utility) does not result in a higher value for $FI$. This property guarantees that the proposed PPaaS framework maintains a good balance between privacy, MGR, and utility.

\subsection{Privacy Metric (Privacy Quantification)}
During the application of each privacy preservation algorithm, the privacy will be quantified empirically using a multi-column privacy metric, considering that the input datasets are n-dimensional matrices. In the proposed setting, we assume that all the attributes of a particular dataset are equally important, and we ensure it by applying z-score normalization to the input datasets. Then we calculate the differential entropy between the perturbed and non-perturbed attributes of the datasets. The correlation of the data distributions of original data and reconstructed data can be effectively used to extract private information by guessing original data with a higher level of accuracy. Consequently, it is essential to take the data's underlying distribution into account when quantifying the inherent privacy~\cite{agrawal2001design}. Differential entropy of a random variable provides an effective mechanism to quantify privacy by considering such side-information into account. The differential entropy $h(A)$ of a random variable A is defined as given in Equation \ref{eqint}. $h(A)$ can effectively be used to measure the privacy of a random variable~\cite{agrawal2000privacy}. $\Omega_A$ denotes the domain of A. $h(A)$ measures the uncertainty inherent in the value of A. $2^{h(A)}$ is proposed to measure the privacy inherent in the random variable $A$. This value (refer to Equation \ref{atpriv}) is also denoted by $\prod(A)$, where $f_A(a)$ is the density function of $A$.

\begin{equation}
\prod(A)=2^{h(A)}
\label{atpriv}
\end{equation}

\begin{equation}
h(A)=-\int_{\Omega_A} f_A(a) log_2 f_A(a) da
\label{eqint}
\end{equation}

Given a random variable $B$, the conditional differential entropy of $A$ is defined according to Equation \ref{eqint2}.

\begin{equation}
h(A|B)=-\int_{\Omega_{A,B}} f_{A,B}(a,b) log_2 f_{A|B=b}(a) da db
\label{eqint2}
\end{equation}

Therefore, $\prod(A|B)=2^{h(A|B)}$ denotes the average conditional privacy of $A$ given $B$. We can use $\prod(A|B)$ to investigate the privacy of an attribute (a data series) after the perturbed version of that attribute is released to a third party. Therefore, the conditional privacy loss of $A$, given $B$, $P(A|B)$ can be given according to Equation \ref{privacyloss}. $I(A;B)$ is known as the mutual information between the random variables $A$ and $B$, where $I(A;B)$ is given in Equation \ref{mutualinfo}. 

\begin{equation}
I(A;B)=h(A)-h(A|B)=h(B)-h(B|A)
\label{mutualinfo}
\end{equation}

Let's consider $A$ to be the original variable and $B$ to be the perturbed version $A$. $P(A|B)$ provides the fraction of privacy of A which is lost by revealing $B$. 

\begin{equation}
P(A|B)=1-\prod(A|B)/\prod(A)=1-2^{h(A|B)}/2^{h(A)}=1-2^{-I(A;B)}
\label{privacyloss}
\end{equation}

Assume that, $B = A + N$, where $N$ represents the noise variable, which is independent of $A$. Since, $A$ and $N$ are independent, $h(B|A) = h(N)$. Consequently, we can represent $P(A|B)$ using Equation \ref{updatedpab} as $I(A;B)=h(B)-h(N)$.

\begin{equation}
P(A|B)=1-2^{-(h(B)-h(N))}
\label{updatedpab}
\end{equation}

From Equation \ref{privacyloss}, $\prod(A|B)$ (the privacy of $A$ after revealing $B$) can be obtained using Equation \ref{finalpriv}.

\begin{equation}
\prod(A|B) = \prod(A) \times (1-P(A|B))
\label{finalpriv}
\end{equation}

The lower the value of $\prod(A|B)$, the lower the privacy of $A$, when $B$ is released. Hence, for a given dataset we consider the minimum of  $\prod(X_i|X_i^p)$ returned by all the attributes to identify the minimum privacy guarantee (where $X_i^p$ represents the perturbed version of the attribute, $X_i)$. To obtain $\prod(X_i|X_i^p)$, we should know the density function of $X_i$. For this purpose, we used an approach which declares a certain number of bins within the range of $0$ to $1$. Next, we assign the values of a particular variable to each bin and find the probability of each bin using the number of values assigned, as shown in Algorithm \ref{attributeinun}, which is used to generate the inherent uncertainty of a particular attribute ($h(attribute)$).

\begin{center}
    \scalebox{0.9}{
    \begin{minipage}{1.1\linewidth}
     \removelatexerror
      \begin{algorithm}[H]
\caption{Generating the inherent uncertainty $h(X)$ of an attribute ($X(attribute)$)}\label{attributeinun}
\linespread{1.5}

\KwIn{
\begin{tabular}{ l c l } 
		$X               $ & $\gets $ & attribute\\
				$bw               $ & $\gets $ & bin window size (default: 0.01)\\
\end{tabular}
}

\KwOut{
\begin{tabular}{ l c l } 
		$h(X)$ & $ \gets $ & the inherent uncertainty of $X$ 
\end{tabular}
}
   
  declare $bn$ (bins) from $0$ to $1$ with an window interval of $bw$\;
  normalize $X$ between $0$ to $1$\;
  assign the values of $X$ to the bins in $bn$\;
  count the number of values assigned to each bin in $bn$\;
  generate the density function of X ($f_X(x)$) by calculating the bin probabilities of each bin in $bn$\;
  $h(A)=-\int_{\Omega_A} f_A(a) log_2 f_A(a) da$\;
  return $h(A)$
  
\end{algorithm}
    \end{minipage}%
    }
     
\end{center}

We use Algorithm \ref{priveval} to generate the minimum privacy guarantee of a perturbed dataset.

\begin{center}
    \scalebox{0.9}{
    \begin{minipage}{1.1\linewidth}
     \removelatexerror
      \begin{algorithm}[H]
\caption{Generating minimum empirical privacy guarantee $min \prod(X_i|X^p_i)$ for a perturbed dataset}\label{priveval}
\linespread{1.5}

\KwIn{
\begin{tabular}{ l c l } 
		$D               $ & $\gets $ & original dataset with $n$ number of attributes\\
		$D^p               $ & $\gets $ & a perturbed instance of the original dataset, $D$\\
				$bw               $ & $\gets $ & bin window size\\
\end{tabular}
}

\KwOut{
\begin{tabular}{ l c l } 
		$min\{\prod(X_i|X^p_i)\}_{i={1}}^n$ & $ \gets $ & the minimum empirical privacy guarantee 
\end{tabular}
}
   \For{each attribute $X_i$ and its perturbed attribute, $X^p_i$}{
    $noise_{X_i}$ = $X^p_i - X_i$\;
    $h(X_i) = Algorithm$\ref{attributeinun}($X_i$)\;
    $h(X^p_i) = Algorithm$\ref{attributeinun}($X^p_i$)\;
    $h(noise_{X_i}) = Algorithm$\ref{attributeinun}($noise_{X_i}$)\;
    $I(X_i;X^p_i) = h(X^p_i) -  h(noise_{X_i})$ \;
    $P(X_i|X^p_i)=1-2^{(-I(X_i;X^p_i))}$\;
    $\prod(X_i) = 2^{h(X_i)}$\;
    $\prod(X_i|X^p_i) = \prod(X_i) \times (1- P(X_i|X^p_i))$\;
   }
   return $min\{\prod(X_i|X^p_i)\}_{i={1}}^n$

\end{algorithm}
    \end{minipage}%
    }
     
\end{center}

\subsection{Attack Resistance Quantification}
\label{priquant}

During the attack resistance quantification, PPaaS runs the corresponding pool of data reconstruction attacks on the perturbed instances. For example, if the pool of perturbation algorithms contain $j$ number of perturbation algorithms, and the pool of data reconstruction attacks contain $k$ number of approaches, testing all $k$ attacks on $j$ perturbation instances of input dataset will produce $j\times k$ number of reconstructed data instances for a given dataset. In the proposed setting, we assume that all the attributes of a particular dataset are equally important, and we make it sure by applying z-score normalization to the input datasets.  After generating each reconstructed data instances, we measure the variance, $V(P)$ (where $P=(X^r-X)$) between the attributes of the corresponding reconstructed data instance and the original dataset. The more different the reconstructed data from original data, the better the perturbation has been. $Var(P)$ provides an effective mechanism to capture this notion ~\cite{chen2005random}. Hence, the higher the $Var(P)$, the higher the difficulty in reconstructing original data from perturbed data. If $\ X^r$ is a reconstructed data series of attribute $\ X $, the level of strength of the perturbation method can be measured using $\ Var(P)$, where $P=(X^r-X)$.  $\ Var(P)$ can be given by Equation \ref{varp}. 

\begin{equation}
Var(P)=Var(p_1,p_2,\dots,p_n)={\frac{1}{n}}\displaystyle\sum_{i=1}^{n}(p_i-\bar{p})^2
\label{varp}
\end{equation}

Next, the attribute having the minimum of all $Var(P)$ (hence the minimum difference between the corresponding attribute) is considered as the most vulnerable attribute of the dataset (or the most successfully reconstructed attribute).  The higher the $Var(P)$, the higher the strength of the corresponding attribute, as $Var(P)$ indicates the difficulty of estimating the original data from the perturbed data~(\cite{chamikara2019efficient}). Equation \ref{minvar} shows the generation of the minimum variance, $Var(P)^{min}$) for a particular reconstructed dataset instance. 

\begin{equation}
Var(P)^{min}=min \{ Var(P_1), Var(P_2), \dots Var(P_n) \}
\label{minvar}
\end{equation}

In this way, PPaaS will produce $t$ number of $Var(P)_{min}$ values if $t$ number of data reconstruction attacks are being tested on a single perturbed instance of the input dataset. From these $t$ instances, we select the minimum variances $Var(P)_{min}$ value, which represents the minimum guarantee of attack resistance of a particular perturbed data instance, as shown in Equation \ref{minvarp}. 

\begin{equation}
Var(P)_{min}=min \{ Var(P)_1^{min}, Var(P)_2^{min}, \dots Var(P)_t^{min} \}
\label{minvarp}
\end{equation}

Finally, we scale the $Var(P)_{min}$ values within 0 and 1, by applying Equation \ref{znorm} to the corresponding pool. The value returned from Equation \ref{znorm} is considered as the input to the FIS (which accepts inputs of range: $[0,1]$).

\begin{equation}
resistance\_input=\frac{Var(P)^i_{min}}{max \{Var(P)^1_{min},\dots, Var(P)^n_{min}\}}
\label{znorm}
\end{equation}

\subsection{Utility Quantification}
\label{utilityquantificaiton}

The accuracy of the results produced by the requested service is evaluated experimentally to generate the empirical utility. If the application being examined is classification, the classification accuracy is generated for all the privacy preservation algorithms in the pool for the corresponding type of data classification. However, if the corresponding pool of applications contains more than one application to be tested, the minimum accuracy (the minimum guarantee of utility) returned by the corresponding perturbed data instance is considered.  

%All the accuracy (utility) values are scaled between 0, and 1 as the range of inputs accepted by the FIS is bounded by the window of $[0,1]$.

For the experimental evaluation of PPaaS we consider only data classification. The utility of data classification can be quantified based on different metrics such as precision, recall, F-measure, accuracy~\cite{sokolova2006beyond}. Any one of these metrics should provide a reasonable approach to measure the utility of a data classification result. It is the application that determines which one of these is the most suitable metric. PPaaS chooses the best perturbed dataset by considering all privacy preservation approaches' relative performance on an input dataset. Hence, the primary requirement of PPaaS is to use only one suitable metric for the utility quantification. For the experimental analysis of PPaaS, we chose classification accuracy measured using Equation \ref{accuracyeq} (where $TP$ = the number of true positives, $TN$ = the number of true negatives, $FP$ = the number of false positives, $FN$ = the number of false negatives) for the utility quantification of the privacy preservation approaches.

\begin{equation}
Accuracy = \frac{(TP+TN)}{(TP+FP+FN+TN)}
\label{accuracyeq}
\end{equation}

\subsection{Algorithm for generating FI}
Algorithm \ref{ppaasalgo} is used for generating $FI$ for a particular pool of privacy preservation algorithms.

\begin{center}
    \scalebox{0.9}{
    \begin{minipage}{1.1\linewidth}
     \removelatexerror
      \begin{algorithm}[H]
\caption{Algorithm for generating $FI$ for a pool of algorithms}\label{ppaasalgo}
\linespread{1.5}

\KwIn{
\begin{tabular}{ l c l } 
		$D               $ & $\gets $ & input dataset\\
				$[pp_1, pp_2,\dots, pp_n]               $ & $\gets $ & pool of privacy algorithms\\
\end{tabular}
}

\KwOut{
\begin{tabular}{ l c l } 
		$BD_i $ & $ \gets $ & selected perturbed dataset \\
		$pp_i $ & $ \gets $ & selected privacy preserving algorithm\\
\end{tabular}
}
   
  perturb $D$ using the pool of algorithms to generate $D_1^p, D_2^p, \dots, D_n^p$\;
  \For{each perturbed dataset, $D_i^p$ }{
  generate minimum privacy guarantee ($pi_i$) using Algorithm \ref{priveval}\;
  generate minimum attack resistance guarantee ($vp_i$) using Equation \ref{znorm}\;
  generate minimum utility guarantee ($u_i$) by running the corresponding application pool on $D_i^p$ (refer to Section \ref{utilityquantificaiton})\; 
  generate the fuzzy index ($FI_i$) by considering $pi_i,vp_i$ and $u_i$, as inputs to the fuzzy inference system ($FIS$)\;
  }
  select the dataset ($BD_i$) that returns the highest $FI$\; 
\end{algorithm}
    \end{minipage}%
    }
     
\end{center}

\section{Results}

In this section, we provide the results of PPaaS in selecting the best perturbed dataset from a particular pool of algorithms. During the experiments, we consider five classification algorithms: Multilayer perceptron (MLP), k-nearest neighbor (IBK), Sequential Minimal Optimization (SVM), Naive Bayes, and J48 ~(\cite{witten2016data}). We use four privacy preservation algorithms: rotation perturbation (RP), geometric perturbation (GP), PABIDOT, and SEAL~(\cite{chamikara2019efficient}), which are benchmarked for utility for the selected classification algorithms~(\cite{chamikara2019efficient}). The algorithms were tested on five different datasets retrieved from the UCI machine learning data repository\footnote{http://archive.ics.uci.edu/ml/index.php}. Table \ref{datasettb} provides a summary of the datasets. For the generation of the minimum guarantee to attack resistance, we used three data reconstruction attacks: (1) naive estimation (naive inference), (2) Known I/O attack ~\cite{okkalioglu2015survey}, and (3) ICA (Independent Component Analysis)-based attacks~\cite{okkalioglu2015survey}. The corresponding data reconstruction attacks were run on the perturbed instances, and the standard deviation of the difference between the normalized original data and the reconstructed data for each instance was recorded. For the known I/O attack, we assumed that 10\% of the original data is known to the adversary. We set the number of iterations to 10 for both RP and GP with a noise factor (sigma) of 0.3 (the default setting). During the experiments, we used a noise standard deviation ($\sigma$) of 0.3 for PABIDOT, whereas an $\epsilon$ of 1 was maintained for SEAL.  All the experiments were run on a Windows 7 (Enterprise 64-bit, Build 7601) computer with an Intel(R) i7-4790 (4$^{th}$ generation) CPU (8 cores, 3.60 GHz) and 8GB RAM. 

\begin{table}[H]
\centering

    \caption{A summary of the datasets used for the experiments.}   
    \label{datasettb}
\resizebox{0.9\columnwidth}{!}{
    \begin{small}
    	\setlength\tabcolsep{5pt} 
        \resizebox{1\columnwidth}{!}{
    \begin{tabular}{l l l l l }
    \hline
{\bfseries Dataset} & {\bfseries  Abbreviation}         & {\bfseries Number of Records}     & {\bfseries Number of Attributes }   &   \bfseries{Number of Classes}   \\
    \hline
Wholesale customers\tablefootnote{https://archive.ics.uci.edu/ml/datasets/Wholesale+customers}       &	 WCDS & 440 \ & 8 \ & 2 	\\
    Wine Quality\tablefootnote{https://archive.ics.uci.edu/ml/datasets/Wine+Quality}      & WQDS & 4898 \ & 12 \ & 7 \\
     Page Blocks Classification \tablefootnote{https://archive.ics.uci.edu/ml/datasets/Page+Blocks+Classification}          & PBDS  & 5473 \ &  11 \ &   5  \\
Letter Recognition\tablefootnote{https://archive.ics.uci.edu/ml/datasets/Letter+Recognition}             &  LRDS	& 20000 & 17 & 26\\

      Statlog (Shuttle)\tablefootnote{https://archive.ics.uci.edu/ml/datasets/Statlog+\%28Shuttle\%29}        &  SSDS & 58000 \ & 9 \ & 7 \\
      HEPMASS\tablefootnote{https://archive.ics.uci.edu/ml/datasets/HEPMASS\#}      & HPDS & 3310816  \ & 28 \ & 2 \\
HIGGS\tablefootnote{https://archive.ics.uci.edu/ml/datasets/HIGGS\#}      & HIDS & 11000000  \ & 28 \ & 2 \\

    \hline
    \end{tabular}
    }
    \end{small} 
    }
\end{table}

In the proposed experimental setting, we consider 25 case studies where each case study considers one of the five classification algorithms and one of the five datasets. We consider a pool of four data perturbation algorithms: RP, GP, PABIDOT, and SEAL;  (CS stands for "case study") as shown in Tables \ref{classyaccuracy} and \ref{rankresults}. Next, we evaluated the performance of each privacy preservation algorithm in each case to generate the ranks (Fuzzy Indices: FIs) and recorded them in Table \ref{rankresults}.

\begin{table}[H]
  \caption{Classification accuracies returned by four privacy-preserving algorithms and five different classification algorithms, and the minimum privacy guarantees generated according to Equations \ref{minvar} and \ref{znorm} using the differences between original and perturbed data. (CS: case study)}  
   \label{classyaccuracy}
    \centering 
    \resizebox{0.8\columnwidth}{!}{
    \begin{tabular}{rllllll | l l}
    \toprule
\multicolumn{1}{l}{\multirow{1}{*}{\textbf{Dataset}}} & \multicolumn{1}{l}{\multirow{1}{*}{\begin{tabular}[c]{@{}l@{}}\textbf{Privacy}\\\textbf{preserving}\\ \textbf{algorithm}\end{tabular}}} & \multicolumn{5}{c|}{\textbf{Utility after privacy preservation}}     & \multicolumn{2}{c}{\textbf{Privacy guarantee}}                         \\ \cline{3-9} 
\multicolumn{1}{l}{}                         & \multicolumn{1}{l}{}  & \multicolumn{1}{l|}{\begin{tabular}[c]{@{}l@{}}\textbf{MLP}\\ \textbf{CS 1}\end{tabular}} & \multicolumn{1}{l|}{\begin{tabular}[c]{@{}l@{}}\textbf{IBK}\\ \textbf{CS 2}\end{tabular}} & \multicolumn{1}{l|}{\begin{tabular}[c]{@{}l@{}}\textbf{SVM}\\ \textbf{CS 3}\end{tabular}} & \multicolumn{1}{l|}{\begin{tabular}[c]{@{}l@{}}\textbf{Naive Bayes}\\ \textbf{CS 4}\end{tabular}} & \multicolumn{1}{l|}{\begin{tabular}[c]{@{}l@{}}\textbf{J48}\\ \textbf{CS 5}\end{tabular}} & \multicolumn{1}{l|}{\textbf{$min(\prod(X_i|X^p_i))_{i={1}}^n$ }} & \multicolumn{1}{l}{\begin{tabular}[c]{@{}l@{}}\textbf{Scaled}\\ \textbf{$min(\prod(X_i|X^p_i))_{i={1}}^n$}\end{tabular}} \\   \hline
    \multicolumn{1}{l}{LRDS}
          & RP    & 0.7404 & 0.8719 & 0.7107 & 0.4841 & 0.6489 & 1.0160 & 0.9981\\
          & GP    & 0.7912 & 0.9305 & 0.7792 & 0.5989 & 0.7054 & 1.0169  & 0.9990\\
          & PABIDOT & 0.7822 & 0.9224	& 0.7848 &	0.6280 &	0.7262 & 1.0179 & 1.0000\\
          & SEAL &  0.8059 &	0.9367 &	0.8171 &	0.6310 &	0.8528 & 1.0157 & 0.9978 \\
          \hline
    \multicolumn{1}{l}{PBDS} 
          & RP    & 0.9200 & 0.9552 & 0.8999 & 0.3576 & 0.9561 & 0.9988 & 0.9979 \\
          & GP    & 0.9024 & 0.9567 & 0.8993 & 0.4310 & 0.9549 & 1.0009 &  1.0000\\
          & PABIDOT & 0.9583 &	0.9476 & 0.9209 & 0.8968 &	0.9492 & 0.9927 & 0.9838\\
          & SEAL &  0.9634 & 0.9673 &	0.9559 & 0.8697 &	0.9634 & 0.9974 & 0.9965\\
          \hline
    \multicolumn{1}{l}{SSDS} 
          & RP    & 0.9626 & 0.9980 & 0.8821 & 0.6904 & 0.9951  & 0.9991 & 0.9992\\
          & GP    & 0.9873 & 0.9981 & 0.7841 & 0.7918 & 0.9959  & 0.9999 & 1.0000\\
          & PABIDOT & 0.9865 &	0.9867 &	0.9280 &	0.9134 &	0.9874 & 0.9920 & 0.9921\\
          & SEAL &  0.9970 &	0.9921 &	0.9851 &	0.8994 &	0.9987 & 0.9961 & 0.9962\\
          \hline
    \multicolumn{1}{l}{WCDS} 
          & RP    & 0.8909 & 0.8500 & 0.8227 & 0.8455 & 0.8682   & 1.0078 & 0.9974\\
          & GP    & 0.9182 & 0.8659 & 0.8500 & 0.8432 & 0.8886   & 1.0078 & 0.9974\\
          & PABIDOT & 0.9045 &	0.8545 &	0.8841 &	0.8886 &	0.8841 & 1.0104 & 1.0000\\
          & SEAL &  0.8932 &	0.8682 &	0.8909 &	0.8841 &	0.8659 & 1.0072 & 0.9968\\
          \hline
    \multicolumn{1}{l}{WQDS} 
          & RP    & 0.4765 & 0.5329 & 0.4488 & 0.3232 & 0.4553  & 1.0268 & 1.0000 \\
          & GP    & 0.4886 & 0.5688 & 0.4488 & 0.3216 & 0.4643  & 1.0267 & 0.9999\\
          & PABIDOT & 0.5412 &	0.6182 &	0.5147 &	0.4657 &	0.4916 & 1.0225 & 0.9958\\
          & SEAL &  0.5392 &	0.6402 &	0.5202 &	0.4783 &	0.8415 & 1.0255  & 0.9958\\
    \bottomrule
    \end{tabular}%
    }
\end{table}%

Table \ref{classyaccuracy} shows the classification accuracy and the minimum privacy guarantee produced for each pool of privacy preservation algorithms. During the  minimum privacy guarantee generation, we used a bin size of 0.01 (the default value) in Algorithm \ref{attributeinun}. In each pool, the input datasets were perturbed using the four privacy preservation algorithms. Then the perturbed data were analysed by each classification algorithm to generate classification accuracy (utility) values. Table \ref{attackresistancevalues}, includes the $\sqrt{min(Var(P))}$ values generated during the attack resistance analysis.  
	
% \end{table}

\begin{table}[H]
  \caption{Analysis on the minimum attack resistance guarantee.}  
   \label{attackresistancevalues}
    \centering 
    \resizebox{0.6\columnwidth}{!}{
    \begin{tabular}{rllllll}
    \toprule
   \multicolumn{1}{l}{\multirow{1}{*}{\textbf{Dataset}}} & \multicolumn{1}{l}{\multirow{2}{*}{\textbf{\begin{tabular}[c]{@{}l@{}}Privacy-\\preserving\\ algorithm\end{tabular}}}} & \multicolumn{5}{c}{\textbf{$\sqrt{Var(P)_{min}}$ values returned under each attack}}                                        \\ \cline{3-7} 
\multicolumn{1}{l}{}                                  & \multicolumn{1}{l}{}                                               & \multicolumn{1}{l|}{\textbf{\begin{tabular}[c]{@{}l@{}}NI\\  \end{tabular}}} & \multicolumn{1}{l|}{\textbf{\begin{tabular}[c]{@{}l@{}}ICA\\ \end{tabular}}} & \multicolumn{1}{l|}{\textbf{\begin{tabular}[c]{@{}l@{}}I/O\\ \end{tabular}}} & \multicolumn{1}{l|}{\textbf{\begin{tabular}[c]{@{}l@{}}$\sqrt{Var(P)_{min}}$\\ \end{tabular}}} & \multicolumn{1}{l}{\textbf{\begin{tabular}[c]{@{}l@{}}scaled\\ $\sqrt{Var(P)_{min}}$ \end{tabular}}} \\ 

    \midrule
    \multicolumn{1}{l}{LRDS}
          & RP    & 0.8750 & 0.4057 & 0.0945& 0.0945 & 0.1353 \\ 
          & GP    & 1.3248 & 0.6402 & 0.0584   & 0.0584  & 0.0836\\
          & PABIDOT & 1.4046 & 0.7038 & 0.6982 & 0.6982  & 0.9994\\
          & SEAL & 1.4061  & 0.7024  & 0.6986 & 0.6986 & 1.0000	\\
          \hline
    \multicolumn{1}{l}{PBDS} 
          & RP    &0.7261 &0.5560 &0.0001  & 0.0001  & 1.4426e-04 \\
          & GP    &0.2845 &0.1525 &0.0000 & 0.0000  & 0.0000 \\
          & PABIDOT &1.4102 & 0.6951 & 0.6755 &  0.6755 & 0.9745\\
          & SEAL &1.3900  & 0.7008  & 0.6932 & 0.6932  & 1.0000\\
          \hline
    \multicolumn{1}{l}{SSDS} 
          & RP    &1.2820 &0.1751 &0.0021  & 0.0021 & 0.0030 \\
          & GP    & 1.4490 &0.0062 &0.0011 & 0.0011 & 0.0016  \\
          & PABIDOT & 1.4058 &0.7069 &0.7031 & 0.7031  & 1.0000 \\
          & SEAL & 1.4065  & 0.7038 & 0.7027 &  0.7027	& 0.9994\\
          \hline
    \multicolumn{1}{l}{WCDS} 
          & RP    &1.0105 &0.6315 &0.0000 & 0.0000  & 0.0000 \\
          & GP    & 1.4620 &0.1069 &0.0000 & 0.0000 & 0.0000\\
          & PABIDOT &1.3680 &0.6771 &0.6512 & 0.6512 & 0.9931\\
          & SEAL & 1.3130 & 0.6775 & 0.6557 &  0.6557  & 1.0000\\
          \hline
    \multicolumn{1}{l}{WQDS} 
          & RP    &1.2014 &0.4880 &0.0057 & 0.0057  & 0.0083 \\
          & GP    & 1.3463 &0.3630 &0.0039 & 0.0039 & 0.0057\\
          & PABIDOT &1.4019 &0.7034 &0.6901 & 0.6901	 & 1.0000	\\
          & SEAL &1.3834  & 0.7018  & 0.6859 & 0.6859	 &  0.9939	\\
    \bottomrule
    \end{tabular}%
    }
\end{table}%

The values in Tables \ref{classyaccuracy} and Table \ref{attackresistancevalues} are evaluated using the proposed fuzzy model to generate the ranks for each privacy preservation algorithm and perturbed dataset as given in Table \ref{rankresults}. The highest ranks generated in each pool of algorithms are in bold and highlighted in colour. Although SEAL has the best performance results in many cases, the table clearly shows that the input dataset and the choice of application (e.g. classification) are also important when  selecting the best privacy preservation approach. Consequently, this result does not mean that SEAL will outsmart other algorithms in other applications with other datasets. 

\begin{table}[H]
  \caption{The best choice of perturbation in each pool based on the highest $FI$ rank values returned.}  
   \label{rankresults}
    \centering 
    \resizebox{0.6\columnwidth}{!}{
    \begin{tabular}{rllllll}
    \toprule
   \multicolumn{1}{l}{\multirow{1}{*}{\textbf{Dataset}}} & \multicolumn{1}{l}{\multirow{1}{*}{\textbf{\begin{tabular}[c]{@{}l@{}}Privacy-\\preserving\\ algorithm\end{tabular}}}} & \multicolumn{5}{c}{\textbf{$FI$ rank values returned in each Case Study (CS)}}                                        \\ \cline{3-7} 
\multicolumn{1}{l}{}                                  & \multicolumn{1}{l}{}                                               & \multicolumn{1}{l|}{\textbf{\begin{tabular}[c]{@{}l@{}}MLP\\ CS 1\end{tabular}}} & \multicolumn{1}{l|}{\textbf{\begin{tabular}[c]{@{}l@{}}IBK\\ CS 2\end{tabular}}} & \multicolumn{1}{l|}{\textbf{\begin{tabular}[c]{@{}l@{}}SVM\\ CS 3\end{tabular}}} & \multicolumn{1}{l|}{\textbf{\begin{tabular}[c]{@{}l@{}}Naive Bayes\\ CS4\end{tabular}}} & \multicolumn{1}{l}{\textbf{\begin{tabular}[c]{@{}l@{}}J48\\ CS 5\end{tabular}}} \\ 
    \midrule
    \multicolumn{1}{l}{LRDS}
          & RP    & 0.2744 & 0.2722 & 0.2744  & 0.2523  & 0.2744 \\
          & GP    & 0.2023 & 0.2005  & 0.2023  & 0.2023  & 0.2023 \\
          & PABIDOT & 0.8104 &  0.8471	& 0.8115  & \cellcolor{blue!25}\textbf{0.8390} & 0.8068\\
          & SEAL & \cellcolor{blue!25}\textbf{0.8190}  & \cellcolor{blue!25}\textbf{0.8479}	 &	\cellcolor{blue!25}\textbf{0.8230} & 0.8378	 &	\cellcolor{blue!25}\textbf{0.8338}\\
          \hline
    \multicolumn{1}{l}{PBDS} 
          & RP    & 0.1496 & 0.1496 & 0.1496 & 0.1496  & 0.1496  \\
          & GP    & 0.1495 & 0.1495 & 0.1495 & 0.1495 & 0.1495\\
          & PABIDOT & 0.8407 & 0.8402 & 0.8378 & 0.8343 & 0.8403\\
          & SEAL & \cellcolor{green!25}\textbf{0.8490} & \cellcolor{green!25}\textbf{0.8491} & \cellcolor{green!25}\textbf{0.8487} & \cellcolor{green!25}\textbf{0.8375} & \cellcolor{green!25}\textbf{0.8490}\\
          \hline
    \multicolumn{1}{l}{SSDS} 
          & RP    & 0.1505 & 0.1505 & 0.1505 & 0.1505 & 0.1505 \\
          & GP    & 0.1500  & 0.1500  & 0.1500  & 0.1500  & 0.1500   \\
          & PABIDOT & 0.8479 & 0.8479 &	0.8453 & \cellcolor{orange!25}\textbf{0.8437} & 0.8479\\
          & SEAL & \cellcolor{orange!25}\textbf{0.8495} & \cellcolor{orange!25}\textbf{0.8493}	 & \cellcolor{orange!25}\textbf{0.8492} & 0.8431 & \cellcolor{orange!25}\textbf{0.8501} \\
          \hline
    \multicolumn{1}{l}{WCDS} 
          & RP    & 0.1495 & 0.1495 & 0.1495 & 0.1495 & 0.1495 \\
          & GP    & 0.1495 & 0.1495 & 0.1495 & 0.1495 & 0.1495 \\
          & PABIDOT & \cellcolor{red!25}\textbf{0.8428} & 0.8325 &	0.8393 & 0.8402 & \cellcolor{red!25}\textbf{0.8393}\\
          & SEAL & 0.8422  & \cellcolor{red!25}\textbf{0.8372} & \cellcolor{red!25}\textbf{0.8418} & \cellcolor{red!25}\textbf{0.8406} & 0.8367\\
          \hline
    \multicolumn{1}{l}{WQDS} 
          & RP    & 0.1522 & 0.1523 & 0.1522 & 0.1522 & 0.1522\\
          & GP    & 0.1513 & 0.1514 & 0.1513 & 0.1513 & 0.1513\\
          & PABIDOT & \cellcolor{purple!25}\textbf{0.8486} & \cellcolor{purple!25}\textbf{0.8398} & \cellcolor{purple!25}\textbf{0.8491} & 0.8368 & \cellcolor{purple!25}\textbf{0.8477}	\\
          & SEAL & 0.8480  & 0.8342	 &  0.8485	 &	 \cellcolor{purple!25}\textbf{0.8426} & 0.8292	\\
    \bottomrule
    \end{tabular}%
    }
\end{table}%

\subsection{Scalability of PPaaS}
\label{scalability}
It is important that PPaaS runs in an  high performance computing-based environment, as it involves multiple processing modules and heavy computation. We can identify two modules of PPaaS: (1) Generating perturbed instances and (2) Running the data reconstruction attacks as those needing the most computation. As shown in Figure \ref{mainflow}, PPaaS can run the steps multiple times until the $FI$ value reaches a certain threshold ($FI_T$). The parallel processing capabilities of implementational components, such as PySpark, allow PPaaS to utilize independent processing modules efficiently. It is also essential that the privacy-preservation approaches used in the PPaaS method repository are efficient enough and capable of dealing with high-dimensional data (e.g. big data). Table \ref{scalability} shows the performance (scalability) of the privacy-preservation algorithms when they are applied to high-dimensional data. For this experiment, we used an SGI UV3000 supercomputer, with 64 Intel Haswell 10-core processors, 25MB cache, and 8TB of global shared memory connected by SGI's NUMAlink interconnect.  As shown in Table \ref{scalability}, PBDOT and SEAL perform extremely well compared to RP and GP, and may be the preferred privacy-preservation algorithms under complex scenarios such as the privacy preservation of big data.

\begin{table}[H]
	\centering
	
	\caption{Efficiency of the privacy preservation algorithm used in the experiments when they are introduced to high-dimensional data}     
	\label{scalability}
	\setlength\tabcolsep{5pt} 
	\begin{small}
		
        \resizebox{0.9\columnwidth}{!}{
		\begin{tabular}{ l l l l l l}
			\hline
			{\bfseries Dataset} & {\bfseries  Dimensions}           & {\bfseries RP  } & {\bfseries GP  } & {\bfseries PABIDOT } & {\bfseries SEAL (ws = 10,000)} \\
		
			\hline
			HPDS        & 3310816$\times$28 & Not converged   & Not  converged 	& 2.9 hours & 97.82 seconds \\
			
			 &	  & for 100 hours & for 100 hours &\\
             HIDS        & 11000000$\times$28 & Not converged   & Not  converged 	& 11.16 hours &  1.02E+03 seconds\\
			
			 &	  & for 100 hours & for 100 hours &\\

			\hline
		\end{tabular}
        }
	\end{small} 
\end{table}
\section{Discussion}

In this paper, we proposed a new paradigm named privacy preservation as a service (PPaaS), to improve the process of privacy preservation of a dataset or application, eventually improving the utility of existing and new privacy preservation approaches. The domain of data privacy contains a plethora of different privacy preservation approaches that have been proposed for different types of applications. However, there are still challenges when it comes to identifying the best privacy preservation method for a given dataset and a certain application; in particular, providing the best utility and maintaining privacy at a high level is difficult. Consequently, it is a highly complex process to identify the best possible privacy preservation approach for a particular application. PPaaS provides a solution by introducing a service-oriented framework that collects existing privacy preservation approaches and semantically categorizes them into pools of applications. 
Developers of new privacy preservation algorithms can introduce their methods to the PPaaS framework and add to the corresponding pools of applications. When a data owner/curator wants to apply privacy-preservation to a particular dataset, PPaaS will rank the methods in the relevant pools of applications with respect to the dataset. The ranks are expressed in the form of a Fuzzy Index ($FI$). $FI$ values are generated using a fuzzy inference system that takes three inputs: privacy (the minimum privacy guarantee), attack resistance (the minimum guaranteed attack resistance), and utility (the minimum guaranteed utility). PPaaS quantifies privacy using a metric ($\prod(X_i|X^p_i)$) based on differential entropy of input data and perturbed data. PPaaS considers the concept of minimum privacy guarantee ($min\{\prod(X_i|X^p_i)\}_{i={1}}^n$), where the minimum of  $\prod(X_1|X^p_1)$ to $\prod(X_n|X^p_n)$ is considered. The strength of the weakest attribute in a perturbed dataset is $min\{\prod(X_i|X^p_i)\}_{i={1}}^n$,  and is called the minimum privacy guarantee. The attack resistance minimum guarantee  is measured by testing the strength of the perturbed instances against the corresponding pool of data reconstruction attacks. For this task, each data reconstruction attack will reconstruct a dataset by attacking the perturbed data instance. Each reconstructed dataset is compared with the original dataset to produce $n$ number of $Var(P)$ values, where $P$ represents the difference between an original attribute and its reconstructed attribute, and $n$ is the number of attributes. From the $n$ number of $Var(P)$ values the minimum $Var(P)$ ($Var(P)^{min}$) is selected to represent the most vulnerable attribute under the corresponding attack. From all the reconstructed instances, the minimum $Var(P)^{min}$ ($Var(P)_{min}$) is selected to represent the overall vulnerability of the corresponding perturbed instance under the given set of attacks. The utility is the accuracy measured under the corresponding set of applications in the application pool. For example, when the application is data classification, PPaaS considers classification accuracy as the utility measurement. PPaaS selects the privacy preservation approach or the perturbed dataset that returns the highest $FI$, which represents the case with the best balance between privacy and utility.

We ran experiments with PPaaS using five different datasets, five different classification algorithms, and four different privacy-preservation algorithms that are benchmarked to produce good utility over the corresponding classification algorithms. Our experiments show that the four privacy preservation algorithms are ranked differently based on the application and the input dataset. The highest values of $FI$ indicate the highest privacy, attack resistance, and utility with the best balance between them. After comparing the $FI$ values (available in Table \ref{rankresults}) generated using the values available in Table \ref{classyaccuracy}, we can conclude that $FI$ provides high values, if and only if all utility, privacy, and attack resistance returned by the corresponding method are high. In all other cases, the fuzzy inference system ($FIS$) produces lower values for the $FI$. Hence, $FI$ enables PPaaS to identify the best-perturbed dataset generated by the most suitable privacy preservation algorithm for the corresponding pool of algorithms and for the corresponding input dataset. As described in the introduction (refer to Section \ref{introsection}), selecting the best perturbation approach has to consider several aspects; this research focuses on privacy, attack resistance and utility, and performance is of secondary importance. By imposing a limit on execution time, PPaaS still ensures that computations will be completed in finite time, as shown in Section \ref{scalability}.

\section{Conclusion}

This paper introduced a novel framework named Privacy Preservation as a Service (PPaaS), which tailors privacy preservation to stakeholders' needs. PPaaS reduces the complexity of choosing the best data perturbation algorithm from a large number of privacy preservation algorithms. The ability to apply the best perturbation while preserving enough utility makes PPaaS an excellent solution for big data perturbation. In order to select the best privacy preservation method, PPaaS uses a fuzzy inference system (FIS) that enables PPaaS to generate ranks that are expressed as fuzzy indices for the privacy preservation algorithms applied to a dataset for a given application. The experimental results show that the fuzzy indices are good indicators of a particular privacy preservation algorithm's ability to maintain a good balance between privacy and utility.

\end{document}